\input phyzzx 
\hsize=40pc
%
\catcode`\@=11 
\def\NEWrefmark#1{\step@ver{{\;#1}}}
\catcode`\@=12 
%
\singlespace

\def\define#1#2\par{\def#1{\Ref#1{#2}\edef#1{\noexpand\refmark{#1}}}}
\def\con#1#2\noc{\let\?=\Ref\let\<=\refmark\let\Ref=\REFS
         \let\refmark=\undefined#1\let\Ref=\REFSCON#2
         \let\Ref=\?\let\refmark=\<\refsend}

\let\refmark=\NEWrefmark

\define\fms{D. Friedan, E. Martinec and S. Shenker, {\it Conformal
invariance, supersymmetry and string theory} Nucl. Phys. {\bf B271}
 (1986) 93.}

\define\astashkevich{A. Astashkevich and A. Belopolsky, {\it String center
of mass operator and its effect on BRST cohomology}, Commun. Math. Phys.
{\bf 186} (1997) 109-136, hep-th/9511111.}

\define\belopolskyzwiebach{A. Belopolsky and B. Zwiebach, {\it Who changes
the string coupling}, Nucl. Phys. {\bf B472} (1996) 109, hep-th/9511077 }

\define\wittenossft{E. Witten, {\it Interacting field theory
of open superstrings}, Nucl. Phys. {\bf B276}  (1986) 291.}

\define\horowitz{G. Horowitz, R. C. Myers, and S. Martin, 
{\it BRST cohomology of the superstring at arbitrary ghost number},
Phys. Lett. {\bf B218} (1989) 309.}

\define\bouwknegt{P. Bouwknegt, J. McCarthy, and K. Pilch, {\it
Ground Ring for the 2D NSR string}, Nucl. Phys. {\bf B377} (1992) 541,
hep-th/9112036.}

\define\polch{J. Polchinski, 
{\it Dirichlet Branes and Ramond-Ramond Charges},
Phys. Rev. Lett. {\bf 75} (1995) 4724, hep-th/9510017.}

\define\sigmaa{N. Berkovits, 
{\it A New Sigma Model Action for the
Four-Dimensional Green-Schwarz Heterotic Superstring},
Phys. Lett. {\bf B304} (1993) 249, hep-th/9303025.}

\define\calcu{N. Berkovits, 
{\it Calculation of Green-Schwarz Superstring Amplitudes using
the N=2 Twistor-String Formalism}, 
Nucl. Phys. {\bf B395} (1993) 249, hep-th/9208035.}

\define\ketov{J. Bischoff, S.V. Ketov and O. Lechtenfeld, 
{\it GSO Projection, BRST Cohomology and Picture-Changing in
N=2 String Theory}, 
Nucl. Phys. 
{\bf B438} (1995) 373, hep-th/9406101.}

\define\twist{N. Berkovits, 
{\it The Ten-Dimensional Green-Schwarz Superstring is a
Twisted Neveu-Schwarz-Ramond String}, Nucl. Phys. 
{\bf B420} (1994) 332, hep-th/9308129.}

\define\araf{
I.Ya. Aref'eva,  P.B. Medvedev and A.P. Zubarev,
{\it Background Formalism for Superstring Field Theory}, 
Phys. Lett. {\bf B240} (1990) 356; 
~~ I.Ya. Aref'eva,  P.B. Medvedev and A.P. Zubarev,
{\it New Representation for String Field Solves the Consistency Problem for
Open
Superstring Field}, 
Nucl. Phys. {\bf B341} (1990) 464.} 

\define\gates{
S.J. Gates, Jr. and H. Nishino, 
{\it Does D = 4, N=8 Supergravity Really Know About Heterotic
Strings?}, 
Class. Quant. Grav. 8 (1991) 809; ~~ 
S.J. Gates, Jr. and V.G.J. Rodgers, 
{\it A Truly Crazy Idea About Type IIB Supergravity and Heterotic
Sigma Models}, 
Phys. Lett. {\bf B357} (1995) 552;~~ 
S.J. Gates, Jr and V.G.J. Rodgers, 
{\it Type B / Type O Bosonic String Sigma Models}, 
Phys. Lett. {\bf B405} (1997) 71.} 

\define\sieg{W. Siegel, {\it Boundary Conditions in 
First Quantization}, Int. J. Mod. Phys. {\bf A6} (1991) 3997.} 

\define\siegelsigma{N. Berkovits and W. Siegel, 
{\it Superspace Effective Actions for 4D Compactifications
of Heterotic and Type II Superstrings}, Nucl. Phys. {\bf B462} (1996) 213,
hep-th/9510106.}

\define\siegelnon{N. Berkovits, M. Hatsuda and W. Siegel, 
{\it The Big Picture}, 
Nucl. Phys. {\bf B371} (1991) 434.} 

\define\sken{J. de Boer and K. Skenderis, 
{\it Covariant Computation of the Low-Energy Effective Action of the
Heterotic Superstring}, Nucl. Phys. {\bf B481} (1996) 129,
hep-th/9608078.} 

\define\central{N. Berkovits, 
{\it Ramond-Ramond Central Charges in the Supersymmetry Algebra
of the Superstring}, 
Phys. Rev. Lett. {\bf 79} (1997) 1813, hep-th/9706024.}

\define\extra{N. Berkovits, {\it Extra Dimensions in Superstring Theory},
to appear in Nucl. Phys. B, hep-th/9704109.}

\define\town{P.K. Townsend, private communication.}

\define\howe{M.T. Grisaru, P. Howe, L. Mezincescu,
B. Nilsson and P.K. Townsend, {\it N=2 Superstrings in a 
Supergravity Background}, Phys. Lett. {\bf 162B} (1985) 116.} 

\define\sanj{S. Rangoolam, private communication.}

\define\rrft{N. Berkovits, 
{\it Manifest Electromagneitc Duality in Closed Superstring Field Theory}, 
Phys. Lett. {\bf B388} (1996) 743, hep-th/9607070.} 

\define\thornyost{C. Preitschopf, C. Thorn and S. Yost, {\it Superstring
field theory}, Nucl. Phys. {\bf B337} (1990) 363.}

\define\siegelzwiebach{W. Siegel and B. Zwiebach, {\it Gauge String Fields},
Nucl. Phys. {\bf B263} (1986) 105 }

\define\zwiebachann{B. Zwiebach, {\it Constraints on covariant
theories for closed string fields}, Ann. Phys. {\bf 186} (1988) 111.}

\define\nelson{P. Nelson, {\it Covariant insertion of general
vertex operators}, Phys. Rev. Lett. {\bf 62} (1989) 993;~~
H. S. La and P. Nelson, {\it Effective field equations for fermionic
strings}, Nucl. Phys. {\bf B332} (1990) 83.}

\define\distlernelson{J. Distler and P. Nelson, 
{\it Topological couplings and contact terms
in 2-D field theory}, Comm. Math. Phys. {\bf 138} (1991) 273.}

\define\bergmanzwiebach{O. Bergman and B. Zwiebach {\it  The 
dilaton theorem and closed string backgrounds}, Nucl. Phys. {\bf B441}
(1995) 76; hep-th/94110047.}

\define\zwiebach{B. Zwiebach, {\it Closed string field theory: Quantum
action and the Batalin-Vilkovisky master equation}, Nucl. Phys. {\bf B390} 
(1993) 33, hep-th/9206084.}

\define\belopolsky{A. Belopolsky, {\it Picture changing operators in
supergeometry and superstring theory}, hep-th/9706033; 
 {\it New geometrical approach to superstrings}, hep-th/9703183;
{\it De Rham 
cohomology of the supermanifolds and superstring BRST cohomology}, 
Phys. Lett. {\bf B403} (1997) 47, 
hep-th/9609220.}

\define\yost{S. Yost, {\it Superstring boundary states}, Nucl. Phys.
{\bf B321} (1989) 629.}

\define\figueroa{J. M. Figueroa-O'Farrill and T. Kimura, Commun. Math. Phys.
{\bf 124} (1989) 105}

\define\zuckerman{B. H. Lian and G. J. Zuckerman, {\it BRST cohomology
of the Super-Virasoro algebras}, Commun. Math. Phys. {\bf 125} (1989) 301.}

\define\henneaux{M. Henneaux, {\it BRST cohomology of the fermionic string}
Phys. Lett. {\bf B183} (1987) 59.}

\define\ramgoolamthorlacius{S. Ramgoolam and L. Thorlacius, {\it D-branes and
physical states}, Nucl. Phys. {\bf B483} (1997) 248; hep-th/9607113.}

\define\sagn{M. Bianchi, G. Pradisi and A. Sagnotti, {\it  
Toroidal compactification and symmetry breaking in open string theories}, 
Nucl. Phys. {\bf B376} (1992) 365.}

\define\polchinskicai{J. Polchinski and Y. Cai, {\it Consistency of
open superstring theories}, Nucl. Phys. {\bf B296} (1988) 91.}

\define\fischlersusskind{W. Fischler and L. Susskind, Phys. Lett. {\bf B171}
(1986) 383; {\bf B173} (1986) 262.}

\def\half{{1 \over 2}}

\def\>{{\rangle}}
\def\<{{\langle}}

\def\p{{\partial}}

\def\L{{\Lambda}}

\def\b {{\beta}}

\def\g {{\gamma}}

\def\fracs#1#2{{\textstyle{#1\over#2}}} 
\singlespace
{}~ \hfill \vbox{\hbox{MIT-CTP-2637}
\hbox{HUTP-97/A025}
\hbox{IFT-P.069/97}
\hbox{hep-th/9711087}\hbox{
} }\break
\title{ON THE PICTURE DEPENDENCE OF RAMOND-RAMOND COHOMOLOGY}
\author{Nathan Berkovits \foot{E-mail address: nberkovi@ift.unesp.br.
Supported in part by CNPq fellowship 300256/94-9.}}
\address{Instituto de F\'{\i}sica Te\'orica, Univ. Estadual Paulista\break
Rua Pamplona 145, S\~ao Paulo, SP 01405-900, Brasil}
\author{Barton Zwiebach \foot{E-mail address: zwiebach@irene.mit.edu.
Supported in part by D.O.E.
contract DE-FC02-94ER40818, and a fellowship of the John Simon Guggenheim
Memorial Foundation.}}
\address{Center for Theoretical Physics,\break
LNS, 
and Department of Physics, 
MIT,\break 
Cambridge, Massachusetts 02139, U.S.A.}

\abstract 
{Closed string physical states are  BRST cohomology classes computed
on the space of states annihilated by  $b_0^-$.  Since $b_0^-$ does not  
commute with the operations of picture changing, BRST cohomologies 
at different pictures need not agree.  We show explicitly that
Ramond-Ramond (RR) zero-momentum physical 
states are inequivalent at different pictures,  and prove that non-zero 
momentum physical states are equivalent in all  pictures.  
We find that D-brane states represent BRST classes that are 
nonpolynomial on the superghost zero modes, while RR gauge fields appear
as polynomial BRST classes. We also 
prove that in $x$-cohomology, 
the cohomology where the zero mode of the spatial
coordinates is included,  there is a unique ghost-number one BRST
class responsible for the Green-Schwarz anomaly, and a unique
ghost number minus one BRST class associated with RR charge. }
\endpage

\chapter{Introduction and Summary}

Ramond-Ramond states of the closed superstring play a crucial 
role in strong-weak duality conjectures. 
At zero momentum, these states 
decouple from perturbative amplitudes, but couple non-perturbatively to
D-brane solitons [\polch]. 
Unlike all other bosonic states of the string or 
superstring, the kinetic terms of Ramond-Ramond (RR) 
fields appear without a dilaton dependent 
$e^{-2\phi}$ factor in the spacetime effective action in string gauge.

Despite the importance of these states, their role in superstring
theory is still poorly understood. 
Sigma models in Ramond-Ramond backgrounds have been constructed 
with worldsheet fields of the Type II superstring 
only in the manifestly spacetime-supersymmetric formalism, using either
the Green-Schwarz [\howe]
or the modified Green-Schwarz description [\siegelsigma].\foot
{Sigma models in Ramond-Ramond backgrounds have also been constructed
in [\gates] using worldsheet fields of the heterotic superstring.}
The 
Green-Schwarz sigma model is probably incomplete since it lacks 
a Fradkin-Tseytlin term coupling the spacetime dilaton to worldsheet
curvature[\town]. Although the modified Green-Schwarz sigma model includes
such a Fradkin-Tseytlin term [\sigmaa,\siegelsigma,\sken], 
it is unsuitable for the uncompactified
superstring since it lacks manifest SO(9,1) covariance. 
Superstring field theory actions for
RR fields 
have been constructed only at the free level[\rrft], and
require extra non-minimal worldsheet variables to be added to the usual
Ramond-Neveu-Schwarz (RNS) variables[\sieg,\siegelnon]. 
 It is not clear whether or not
a consistent free closed string field theory can be formulated using only
RNS variables, as would have been naturally expected to be the case. 

The lack of a free RR kinetic term for closed superstrings using the
standard RNS variables  
implies that, precisely speaking, we do not really  
have a definition of 
physical states in the RR sector. 
For bosonic strings, physical closed string states are simply  
defined as BRST cohomology classes computed on the {\it semirelative
complex}, i.e., on the subvector space of states that are annihilated by the
antighost zero mode $(b_0-\bar b_0)$. The necessity to restrict the
set of states of the conformal field theory
to the semirelative complex is well understood and can be  
seen in a variety of ways. \foot{These include the need to accommodate
topological constraints on sections of 
bundles over moduli spaces of Riemann surfaces [\nelson], the
requirement of having an action principle for the
cohomology problem [\siegelzwiebach,\zwiebachann,\zwiebach],
and the required existence of a        
zero-momentum physical (ghost) dilaton state 
[\distlernelson, \bergmanzwiebach].}  
For similar reasons, and a new one (see below), 
it seems clear that 
the restriction to the semirelative complex is also 
necessary
in closed superstring
fIeld theory. 
On the other
hand, BRST cohomology in superstring theory 
requires a choice of picture
for the superghost vacua [\fms]. 
While it was known that {\it absolute} BRST
cohomology, i.e., the cohomology computed on the unrestricted complex,
is independent of the choice of picture [\horowitz,\fms],  
it was not known before this
paper if semirelative cohomology would be independent of such choice. 
Indeed, the literature deals with the 
chiral superstring complex only [\henneaux,
\figueroa,\zuckerman], and the  closed superstring semirelative
cohomology
appears not to have been computed.

One could ask if any additional constraints should
be imposed on the closed superstring fields, 
e.g. perhaps string fields in the RR sector need to be
annihilated by $(\beta_0 -\bar\beta_0)$ where $\beta_0$ and $\bar\beta_0$
are the zero modes of bosonic ghosts. 
If spacetime supersymmetry is to be an off-shell
symmetry, however, powerful constraints on such possibilities
arise.   In the NS-NS sector,
the constraints defining the BRST complex (the off-shell
string field) are
$(b_0 - \bar b_0)\Phi=0$, and
$(L_0 - \bar L_0)\Phi=0$ where $L_0 -\bar L_0 =0$ is the usual
level-matching condition.  Since the spacetime-supersymmetry
generators anti-commute with
$b_0-\bar b_0$ \foot{This is easily seen from the fact
that the supersymmetry charge is independent of the  
reparametrization ghost $c(z)$ in both
the $-\half$ and $+\half$ pictures [\fms].}, and commute with
$(L_0 - \bar L_0)$, it is natural to impose the same constraints
on the R-NS, NS-R, and RR BRST complexes.\foot{Note that in the
closed superstring field theory action constructed using non-minimal
worldsheet variables [\rrft], all sectors of the superstring satisfy 
these off-shell 
constraints.} 
In this way, supersymmetry manifestly maps the complexes into
each other, as should be the case for generators of off-shell
symmetries. 

Since the superghost fields do not have simple
commutation relations with the supersymmetry operators,
further constraints based on the superghosts in the
R-NS, NS-R or RR sectors are likely to imply further
conditions in the NS-NS sector. Moreover, all such conditions
should not affect BRST cohomology at non-zero momentum and at
the physical ghost number, for this cohomology
matches the light-cone physical states.
While all
this argumentation does not prove that further conditions
cannot be imposed, it does not look easy to do.
In this paper, we only impose the $(b_0-\bar b_0)=0$ and
$(L_0 -\bar L_0)=0$ conditions, we show that the non-zero
momentum cohomology is the expected one, and we find that
the subtle zero-momentum cohomology appears to be
physically sensible.

Using the standard RNS description, we    
perform a careful analysis of BRST 
semirelative cohomology in all sectors of closed superstrings,
focusing in particular on the Ramond-Ramond sector.  
We show that
{\it the  Ramond-Ramond BRST semirelative classes
are indeed picture-dependent}. 
The analysis of semi-relative
cohomology is performed at all ghost numbers and in the pictures
$(-1/2, -1/2)$, $(-1/2,-3/2)$,  
$(-3/2, -1/2)$, and $(-3/2,-3/2)$, which are the four 
Ramond-Ramond  pictures
in which all positive oscillator modes of the left and right-moving
$(\beta,\gamma)$ ghosts annihilate the vacuum. 

In section 2 we prove that for every sector of the closed
superstring, the BRST semi-relative
classes at {\it non-zero momentum} are the same 
in all possible pictures. 
This is expected since, at
non-zero momentum, one 
can compute the physical spectrum in light-cone gauge. 
At a more technical level, this happens because at non-zero momentum
one can define both picture-raising and picture-lowering operators
that commute with $(b_0-\bar b_0)$.

In section 3, we show that the RR semirelative cohomology at  
zero momentum is indeed inequivalent
in the four different pictures mentioned above. 
This inequivalence is caused by the fact
that $b_0- \bar b_0$ does not commute with any picture-lowering operator
containing a finite number of terms.\foot{It is interesting to note
that there are similar problems with defining a picture-lowering
operator for the {\it open} 
N=2 string [\calcu]. As shown in [\ketov], 
a BRST-invariant inverse to the N=2
picture-raising operators can be constructed only if one allows 
an infinite number of terms. This suggests that some of the
techniques developed in this paper may also be applicable for 
open  N=2 strings.} 
On the other hand, both for the NS-R (or R-NS) sector of closed superstrings
and for the
R-sector of heterotic strings there are two obvious pictures to 
consider and we find that zero-momentum semirelative states are
picture independent.

In our analysis of zero-momentum  RR states, 
we found it is useful to distinguish between
the conventional 
``finite'' cohomology and a novel ``infinite'' cohomology [\sanj].   
These are
just BRST cohomologies computed in two complexes that are defined
in a slightly different ways. 
Taking basis vectors to be monomials built as the product
of oscillators
acting on the vacuum, states in the
{\it finite complex} are restricted to be 
finite linear combinations of the basis vectors, while 
we impose no such restriction for the {\it infinite complex}. 
The infinite complex
is just an enlarged version of the finite complex, and thus the
usual phenomena can occur;  finite BRST classes might become
trivial in the infinite complex, and there may exist infinite BRST
classes that have no representatives in the finite complex. These
latter states will be called {\it strictly infinite} BRST classes.

We find that the finite and infinite cohomologies at zero momentum
are equivalent 
both in the $(-1/2,-1/2)$ picture and in the  
$(-3/2,-3/2)$ picture. 
This happens because upon restriction to a fixed ghost number and to 
$L_0=\overline L_0 =0$, 
the state space in these pictures 
becomes finite dimensional and the finite and
infinite complexes coincide. 
On the other hand, for the mixed $(-3/2,-1/2)$ picture (or the $(-1/2,-3/2)$
picture),
restriction to a fixed ghost number and to $L_0=\overline L_0 =0$ 
leaves an infinite dimensional state space. As a result, 
the finite and 
infinite cohomologies are different 
in the mixed $(-3/2,-1/2)$ picture; the same is true also for the 
$(-1/2,-3/2)$  picture.
Furthermore, we find in these mixed pictures that 
the finite cohomology classes are all trivial in the
infinite complex, and that there exist strictly infinite BRST classes.
Therefore the infinite cohomology in these mixed
pictures is all strictly infinite,
and it is natural to think of the finite and infinite cohomologies
as disjoint vector spaces.

While it is thus clear that (zero-momentum)
semirelative cohomologies are not
the same for different pictures, we prove some relations that
show that  the $(-1/2,-1/2)$ and the
$(-3/2,-3/2)$ pictures {\it together} contain the information that is
encoded in the mixed picture $(-3/2,-1/2)$ (or the picture
$(-1/2,-3/2)$). 
For zero momentum states, we prove that the finite cohomology at picture
$(-1/2,-3/2)$ or $(-3/2,-1/2)$
is equivalent to the cohomology at picture $(-3/2,-3/2)$,
while the infinite cohomology at picture
$(-1/2,-3/2)$ or $(-3/2,-1/2)$
is equivalent to the cohomology at picture $(-1/2,-1/2)$.
In summary, for Ramond-Ramond zero-momentum semirelative cohomology we have the
following isomorphisms 
$$\matrix{ H_{-{3\over 2}, -{3\over 2}} & \longleftrightarrow &
H^{fin}_{-{3\over 2}, -{1\over 2}} & {}&\cr\cr
& {}& H^{\infty}_{-{3\over 2}, -{1\over 2}}& \longleftrightarrow
& H_{-{1\over 2} -{1\over 2}} } \eqn\relation$$

In addition to being of mathematical interest, the distinction between
finite and infinite cohomology is also of physical interest. Boundary
states for superstring D-branes are shown to be representatives for
strictly infinite BRST classes (classes in 
$H^{\infty}_{-{3\over 2}, -{1\over 2}}$). 
This is a superstring phenomenon.   
Although D-brane states in bosonic string theory are
also nonpolynomial functions of the oscillators
(and thus  
define an infinite cohomology class),
all except a
finite set of basis vectors entering into the
expansion of the state are BRST exact [\ramgoolamthorlacius].  
Therefore, the class of the bosonic
D-brane state has a representative in the finite complex.
As defined above, this means that
the BRST class associated to a bosonic D-brane state,
in contrast to that of superstring D-brane states, is not strictly infinite.
The difference comes from the fact that superstring D-brane states 
have non-polynomial dependence on bosonic ghost zero modes. 
We also show that the finite cohomology contains RR
central charges and zero-momentum RR gauge fields [\central].

In section 4, we refine our analysis of zero-momentum RR
cohomology to
$x_0$-cohomology, where the states in the BRST complex
can explicitly depend on the zero mode of the non-compact
bosonic coordinates $X^\mu (z,\bar z)$ [\astashkevich,\belopolskyzwiebach]. 
We explain why this is the cohomology problem that must be used to discuss
anomalies, and the relation to the Fischler-Susskind mechanism.
In $x_0$-cohomology, ghost-number one states are associated with anomalies, 
ghost-number zero states are associated with zero-momentum
physical fields that cannot be gauged away even with gauge parameters
that diverge at infinity, and ghost-number minus one states are associated 
with global symmetries; for superstrings, super-Poincar\'e 
generators.  

In the Type-I and Type IIB
Ramond-Ramond sector with infinite cohomology, we find
the expected $D_9$-brane anomaly at ghost-number plus one. In the Type IIB
Ramond-Ramond sector with finite cohomology, we find the zero-momentum
axion at ghost-number zero. The zero-momentum axion
has properties quite analogous
to those of the zero-momentum ghost dilaton;  it 
would be BRST trivial had we not imposed
the semirelative condition, and, 
it is also present in the relative closed string cohomology
(the cohomology on the complex 
where states satisfy $b_0 \hskip-2pt= \hskip-2pt\bar b_0 
\hskip-2pt=\hskip-2pt0$).  
Finally, in the Type IIA Ramond-Ramond sector 
with finite cohomology, we find a zero-brane charge at ghost-number minus one. 
The presence of this 
zero-brane charge suggests that an eleventh coordinate 
may be present in superstring theory without
having to introduce membranes [\extra].  

An important question that remains open is the construction 
of a superstring field theory using only RNS variables. Our work here
should help understand the role of the various pictures. Further
geometrical insight into picture changing [\belopolsky] could also be of use. 
We believe our work gives evidence that the intricate
structure of pictures in superstring theory is not a technical nuisance
but rather a key element in the future understanding  of duality
transformations, D-branes, and the non-perturbative physics of 
superstring theory. 

\chapter{ The cohomology at nonzero momentum}

The main purpose of the present section is to show that
at non-zero momentum the closed superstring cohomology,
defined as the BRST cohomology evaluated on the subcomplex
where all states are annihilated by $b_0 - \bar b_0 \equiv b_0^-$,
is the same in all pictures. The result will hold for each sector
of the closed superstring, that is, for the NS-NS sector, for the NS-R
sector, and for the RR sector.

This result  follows from a similar result for the chiral
complex,  the complex associated to the left moving part of
the conformal theory.   
Let ${\cal H}_P$ denote the  relative chiral complex,
{\it i.e.}~the states of the chiral sector
that have picture number 
$P$ and are annihilated by $b_0$.\foot{Note that in defining
the chiral relative complex, we only impose a $b_0=0$ condition.  
This differs from the usual
approach where the relative complex is defined as
one where the computations
of BRST cohomology do not involve ghost zero modes [\zuckerman].} 
 These states may be in the NS or R
sectors. Moreover, let  $H_P(Q,b_0)$ denote the
cohomology of $Q$ on ${\cal H}_P$, {\it i.e.}~the 
{\it relative } cohomology at picture number $P$. We now claim that the
following result is true :  

\medskip
\noindent
{\bf Theorem:} For any fixed non-zero momentum, 
the chiral
relative cohomologies
$H_P (Q,b_0)$ and $H_{P-1}(Q,b_0)$ are isomorphic vector spaces. This holds
both for the NS and for the R sectors.

\noindent
{\bf Proof.} Our strategy to establish this result is similar to that
used in [\horowitz], namely, we  identify an invertible picture
changing operator acting on the relative cohomologies.
As we will show, the required operator is the momentum operator
in the $-1$ picture,  
$$\widetilde p^\mu = \oint  {dz\over 2\pi i}  
e^{-\phi(z)} \psi^\mu(z)\, ,\eqn\one$$
which was used in [\wittenossft] to discuss
supersymmetry charges. 
This operator carries picture number minus one, and thus can be used
as an inverse picture changing operator.  
Since the integrand   
does not involve the conformal ghost $c$,  we have that  
$[b_0, \widetilde p^\mu] = 0$.  
Therefore, 
$\widetilde p^\mu : {\cal H}_P \to {\cal H}_{P-1}$, that is, 
$\widetilde p^\mu$ maps
the relative chiral complexes at different pictures.
Furthermore, given that 
$$[ Q,  e^{-\phi} \psi^\mu (z) ] = 
\partial ( c e^{-\phi} \psi^\mu ) (z) \, , \eqn\two$$
one finds that $[Q , \widetilde p^\mu] = 0$.  
As a consequence, $\widetilde p^\mu: H_P(Q,b_0)\to H_{P-1}(Q,b_0)$. 
It remains to show that $\widetilde p^\mu$ 
has a well defined inverse.  
To this end, consider now the picture changing operator $X_0$ defined as
[\fms]
$$X_0 = \oint {dz\over 2\pi i z} X(z)\,, \eqn\three$$
where the local operator $X(z)$ is given by\foot{Our conventions
can be found in section 3.1.}  
$$\quad  X(z) = \{ Q , \xi (z) \}= c\partial \xi\, 
- \,\fracs12 e^\phi \psi\cdot \partial 
+ (\p\eta) e^{2\phi} b  
+\p(\eta e^{2\phi} b) \,.\eqn\threex$$
This operator is BRST nontrivial despite appearances, 
and it commutes with $Q$. Furthermore, as mentioned
in [\bouwknegt], $X_0$ commutes with the zero mode of the antighost field
$$[b_0 ,  X_0 ] = (\partial \xi )_0 = 0\, . \eqn\four$$
It follows that $X_0\hskip-2pt: H_{P-1}(Q,b_0) \to H_P(Q,b_0)$ 
and is therefore a candidate for an inverse to $\widetilde p^\mu$.   

\medskip
\noindent
We now claim that the following
operator relations hold 
$$\eqalign{
X_0  \, \widetilde p^\mu &= -\fracs12 \,p^\mu + \{ Q ,  m^\mu \} \, , \cr 
\widetilde p^\mu X_0 & = -\fracs12\, p^\mu + \{ Q,n^\mu \}\, ,\cr} \eqn\five$$
where $p^\mu$ is the momentum operator 
$$p^\mu = \oint {dz\over 2\pi i}\, \partial X^\mu \,, \eqn\six$$
and  the operators $m^\mu$ and $n^\nu$ satisfy
$$\eqalign{ 
[\, b_0\, , m^\mu\, ] &= 0\, , \cr
[\, b_0\, , n^\mu\, ]  &= \, 0\, . \cr } \eqn\seven$$
Before establishing these relations let us  prove that they
imply the desired result, namely,
that at any fixed non-zero momentum the relative cohomologies $H_P(Q,b_0)$ and 
$H_{P-1}(Q,b_0)$ are isomorphic. 
For this, it is enough to show that the map  
$\widetilde p^\mu : H_P(Q,b_0) \to H_{P-1}(Q,b_0)$ is
(i) one to one and  (ii) onto. One to one:  
consider two states $|a\rangle $ and $|b\rangle$ of the
same momentum, and both belonging to $H_P(Q,b_0)$. Moreover, assume that
$\widetilde p^\mu \ket{a}  = \widetilde p^\mu \ket{b}$ on $H_{P-1}(Q,b_0)$. 
Explicitly, this means that  
$$\widetilde p^\mu \, ( \ket{a}  - \ket{b}) = Q\ket{\epsilon^\mu
}\,, \eqn\eight$$
where $b_0 \ket{\epsilon^\mu} = 0$. Then multiplying by $X_0$ one finds
$$X_0 \, \widetilde p^\mu \, ( \ket{a}  - \ket{b} )= 
Q X_0 \ket{\epsilon^\mu} \,,\eqn\nine$$
and using the first equation above 
$$-\,\fracs12\, 
p^\mu ( \ket{a}  - \ket{b} ) =   Q \bigl(  -m^\mu  ( \ket{a}  - \ket{b})  -  
X_0 \ket{\epsilon^\mu}\bigr)\,. \eqn\ten$$
Since $p^\mu$ has a definite non-zero eigenvalue in the left hand side, 
and the gauge parameter on the right hand side is annihilated by $b_0$, 
this equation shows
that $\ket{a} = \ket{b}$ on $H_P(Q,b_0)$. 
This shows that the map is one to one.

To show that the map is onto, consider any  nontrivial state 
$\ket{d} \in H_{P-1}(Q,b_0)$
of fixed non-vanishing momentum.  
We claim it is the image under $\widetilde p^\mu$
of the state  
$\ket{d'}  = -2\, X_0 {1\over p^\mu}  \ket{d}$  
where $\mu$ is chosen in a direction for which $p^\mu \ket{d}\neq 0$.
Indeed, it is clear that 
$\ket{d'} \in H_P(Q,b_0)$, and moreover, using the second relation
in \five\ 
we find 
$$\widetilde p^\mu \ket{d'}
= \ket{d} -2 Q {1\over p^\mu} m^\mu \ket{d}\,  \eqn\eleven$$ 
where it is understood that there is no sum over $\mu$ in the second term. 
Since the second term in the right hand side is an 
allowed gauge parameter, this
verifies that the map is onto. 
This completes the verification that  equations 
\five\ imply isomorphic  non-zero momentum 
cohomologies in the various pictures. 

\medskip
\noindent We must now establish equations \five.  
We begin by evaluating $X_0 \,\widetilde p^\mu$, 
$$X_0 \, \widetilde p^\mu =\oint {dz_1\over 2\pi i z_1} \,
X(z_1) \, \oint {dz_2\over 2\pi i}  \,  
(e^{-\phi} \psi^\mu)\,(z_2). \eqn\twelve$$
Using the relation
$$X(z_1 ) - X (z_2')  = \bigl\{ Q , \int_{z_2'}^{z_1}  du 
\partial \xi (u) \, \bigr\}\,, \eqn\diffx$$
we find
$$\eqalign{
X_0 \, \widetilde p^\mu &=  \oint {dz_1\over 2\pi i z_1} 
\oint {dz_2\over 2\pi i}  \, X(z_2') \, 
(e^{-\phi} \psi^\mu)\,(z_2) \cr
&\, +\oint {dz_1\over 2\pi i z_1} 
\bigl\{ Q , \int_{z_2'}^{z_1}  du \,\partial \xi (u) \, \bigr\} \,
\oint {dz_2\over 2\pi i}  \,  (e^{-\phi} \psi^\mu)\,(z_2)\,,\cr} \eqn\tve$$
where we define $z_2' = z_2 + \epsilon$, and we are taking the limit 
as $\epsilon \to 0$.
In the last term of the right hand side, the integral 
to the right can be placed inside
the BRST commutator, and in the first term 
we can evaluate explicitly the operator
product to find   
$$X_0 \, \widetilde p^\mu =  -\fracs12 \,p^\mu  
+ \biggl\{ Q , \oint {dz_1\over 2\pi i z_1} 
\oint {dz_2\over 2\pi i} \int_{z_2}^{z_1}  
du\, \partial \xi (u) \,  (e^{-\phi} \psi^\mu)\,(z_2)\biggr\} \, .\eqn\fift$$
This equation has the form anticipated in \five\ where we can now see 
the explicit
form of the  operator  $m^\mu$, and confirm that it commutes with $b_0$.  
The second equation in \five\ is proven in an exactly identical fashion, 
the operator $n^\mu$ differs
from $m^\mu$ only in the order of the operator product $\partial \xi$ and
$e^{-\phi} \psi^\mu$ in the above expression. This concludes the 
proof of equations \five, and therefore completes the proof of the
theorem.

We note that the proof presented above does not depend on the sector
of the superstring we are considering. For different sectors the picture
numbers $P$ are of different type (integer for NS and 
half integer for R), but no separate
treatment is required, as picture changing
operators satisfy the same identities as they act on the
two different sectors. 

We also note that the above proof does more than establish an
isomorphism of vector spaces. It produces a ``canonical" identification
in the sense that there are natural picture changing operators
implementing the isomorphism. 
This means that we expect states related by the isomorphism to be
physically equivalent.

\goodbreak
\noindent
{\bf Corollary} At fixed nonzero momentum, the closed
superstring semirelative cohomologies
$H_{P,\bar P} (Q, b_0^-)$ of each sector (NS, NS-R, and RR)
are isomorphic for all pictures $P,\bar P$.
\nobreak
\medskip
\noindent
{\bf Proof.} 
The corollary follows by simple observations.
First, once we guarantee that $\widetilde
p^\mu$, $X_0$, $m^\mu$, and  $n^\mu$ commute with $b_0$, being
holomorphic, they will commute with $b_0^- = b_0-\bar b_0$. Moreover
equations \five\ hold for $Q$ the total BRST operator. It thus follows
that $p^\mu$ and $X_0$ guarantee the isomorphism (at fixed momentum)
between
$H_{P,\bar P}(Q,b_0^-)$ and $H_{P-1,\bar P}(Q,b_0^-)$. 
Completely analogous remarks hold for the antiholomorphic sector. 
This completes the verification of the corollary.

\chapter{Cohomology at Zero-Momentum}

In the previous section, we proved that the
semi-relative
cohomology at non-zero momentum of the closed RNS superstring
is isomorphic in all pictures. This proof holds for all sectors of
the superstring, i.e. the NS-NS, NS-R, and R-R sectors.
At zero momentum, however, there is a more complicated
relationship between the semi-relative cohomologies
in different pictures. In this section we will explore the
picture dependence of the zero-momentum cohomology.

We will not attempt here a full analysis of all possible
pictures. We will only relate the
cohomologies for 
superstring pictures in which all positive modes of the
worldsheet fields annihilate the vacuum.\foot{All other pictures 
contain  states with 
arbitrarily negative energy, e.g. $(\g_{1/2})^n |0\>$ in the $0$ picture
of the NS sector. This is not truly pathological
since at any given ghost number the energy is bounded below.
One can certainly work with these pictures, but they have been used
less that the canonical pictures. A superstring field theory with 
NS picture $0$ has been considered in 
[\thornyost,\araf]. } This includes the $-1$
picture in the NS sector (where $\b_r$ and $\g_r$ annihilate
the vacuum for all $r\geq \half$), the $-3/2$ picture in the R sector
(where $\b_{n+1}$ and $\g_{n}$ annihilate the vacuum for $n\geq 0$)
and the $-1/2$ picture in the R sector
(where $\b_{n}$ and $\g_{n+1}$ annihilate the vacuum for $n\geq 0$). 

This section is organized as follows. In subsection 3.1 we give our
conventions. In subsection 3.2 
we show for the chiral Ramond complex that 
the zero-momentum relative cohomology,
i.e., the cohomology in the chiral subcomplex of zero-momentum
states annihilated
by $b_0$, is not the same in pictures $-1/2$ and $-3/2$. This shows that
the theorem proven in section 2 does not extend to zero momentum.

In subsection 3.3, we extend the analysis to the semirelative RR complex
of zero momentum states. We consider the four cases defined by pictures
$(P, \bar P)$, where $P$ and $\bar P$ are either $(-1/2)$ or $(-3/2)$.
We note that when $P\not=\bar P$, the nontrivial physical states would
be BRST trivial if we allowed gauge parameters that are infinite linear
combinations of Fock space states. This suggests refined definitions of
BRST complexes, a finite complex where states must be finite linear
combinations of Fock space states, and an infinite complex where there
is no such condition. We then compute explicitly these two types of
cohomology for the four pictures in question.

In subsection 3.4, we relate the various cohomologies obtaining
the relations indicated
in equation \relation\ of the introduction. To this end, we build
a suitable picture-raising and picture-lowering operator, 
${\cal X}_0$ and ${\cal Y}_0$, that  
map the cohomologies into each other. We observe that the maps
are one-to-one and surjective maps, and thus define canonical 
isomorphisms of the cohomologies. This should imply that these cohomologies
represent the same physical content.  

Finally, in subsection 3.5 we show that the two possible pictures
that can be used to describe the R-sector of heterotic strings yield
equivalent zero momentum semirelative cohomologies. Similarly,   
the two possible pictures
that can be used to describe the R-NS sector of closed superstrings
yield equivalent zero-momentum  semirelative cohomologies. For both
cases, 
we build suitable picture changing operators that establish
the isomorphisms.  

\section{Notation and conventions for vacua}

The chiral Ramond vacuum in the $P$ picture will be denoted by 
$|P\>^\alpha$ where $\alpha$ is a sixteen-component
Majorana-Weyl spinor index. Its vertex operator in
terms of the SL(2)-invariant vacuum $\ket{{\bf 1}}$ is  
$c\hskip1pt S^\alpha e^{P \phi}(z)$
where $c(z)$ is the reparametrization ghost,
$S^\alpha(z)$ is the spin field of conformal weight $5/8$, 
and
$\phi(z)$ is the chiral boson coming from fermionization of the
$(\beta,\gamma)$ system as $\beta = e^{-\phi}\p\xi $
and $\gamma = e^{\phi}\eta$. We will write
$$|P\>^\alpha = c S^\alpha e^{P \phi} (0)\ket{{\bf 1}} \,,\eqn\vacdef$$
and, by construction this state is annihilated by $b_0$ (but not by
$c_0$) and by
all positively moded oscillators of the reparametrization ghosts
$(b,c)$ and the matter fields. 

The inequivalent Majorana-Weyl spinor will be denoted as $|P\>^{\alpha'}$.
The zero modes of the worldsheet fermions act on the vacua as
$$\eqalign{ \psi_0^\mu \ket{P}^\alpha
&= \Gamma^{\mu\,\alpha}_{~~~\beta'}\ket{P}^{\beta'}\,, \cr
\psi_0^\mu \ket{P}^{\beta'}
&= \Gamma^{\mu\,\beta'}_{~~~\alpha}\, \ket{P}^{\alpha}\,, \cr}\eqn\zmactiona$$
and therefore the two vacua have opposite GSO (chirality) eigenvalue.
If we conventionally
declare $\ket{-\fracs12}^\alpha$ to have GSO eigenvalue $+1$,
the GSO-projection implies that states  must be built on vacua 
$\ket{P}$
with spinor index  $\alpha/ \alpha'$ when
$(P + 1/2 + N)$ is even/odd, where  
$N$ is the number of $\b_0$'s plus the number of $\g_0$'s. 
For notational convenience, 
we will leave the spinor index out of many equations. The spinor type can
be deduced from the above rule. Comments made for vacua with the spinor index
left out hold for both kind of spinor indices. 

For convenience, we will set the ghost numbers
of the $|P\>$ vacua to be zero for all $P$. This
follows from defining the ghost-number current as [\twist]   
$$j_{ghost} = -b c + \eta\xi\, ,  \eqn\ghostdef$$ 
rather than as $j_{ghost} = -b c - \partial \phi$, which is the
conventional definition. Note that the ghost numbers of $Q, b,c,\beta$, and
$\gamma$ are unchanged. In this convention, $\xi$ carries ghost number $-1$,
and $\eta$ carries ghost number $+1$. The picture changing operators
$X(z)$ and $Y(z)$ now carry ghost number zero. Both for $P=-1/2$ and
for $P=-3/2$, the $L_0$ eigenvalue of $\ket{P}$ is zero. In these
two pictures, as mentioned earlier, all positively moded superghost oscillators
kill the vacuum. Moreover,  $\beta_0\ket{-\fracs12} =0$ and
$\gamma_0\ket{-\fracs32}=0$. 

The BRST operator in the chiral complex is defined
as 
$$Q = \oint {dz\over 2\pi i}\, \Bigl\{ c (T_M 
+\fracs12T_{gh}) + \gamma (G_M + \fracs12 G_{gh} )\Bigr\}\,, \eqn\qbrst$$
and it is strictly conserved, squares to zero,
and satisfies
$$\eqalign{
\{ Q, b(z)\} &= \,\,T_M + T_{gh}\,,\cr
[ Q , \beta(z) ] &=  G_M(z) + G_{gh}(z)\,,\cr}\eqn\brstprop$$
where\foot{We use the standard ope's $c(z)b(w)\sim
\g(z) \b(w)\sim (z-w)^{-1}$.}
$$\eqalign{
T_M(z) &= -\fracs12 \partial X \cdot\partial X
 -\fracs12 (\partial \psi)\,\cdot \psi\,,\cr
T_{gh}(z) &= c\partial b + 2\partial c b
 -\fracs12 \gamma\partial \beta\, -\,\fracs32 \partial 
\gamma \beta\,, \cr
G_M(z) &= - \fracs12 \psi\cdot \partial X\,,\cr
G_{gh}(z) &= c\partial \beta + \fracs32 \partial c\beta - 2 b\gamma \,.
} \eqn\tgho$$   
In the R-sector, the BRST operator can be expanded as
$$Q = L_0c_0  +  (- \g_0^2 + M)
 b_0 + D_R \g_0 + N \b_0  + \widetilde Q\eqn\expbrst$$  
where $M, N$ and $\widetilde Q$ are independent of ghost and
matter zero modes. Here
$D_R =\fracs12 \, \psi_0^\mu p_\mu + \cdots$, and 
$L_0 = \fracs12 \, p^2 +\cdots$  where the dots indicate
terms independent of ghost and matter zero modes.

In the  R-R sector, states are built on vacua 
of the type $(\ket{P} \otimes
\ket{\bar P})$ where the spinor index on each vacuum is determined
by the GSO projection. The left moving GSO projection is defined
as above, and the right moving  GSO-projection implies that the second 
spinor index is $(\alpha/\alpha')$ when
($\bar P + 1/2 + \bar N +y$) is even/odd, where 
$\bar N$ 
is the total number of $\bar\b_0$'s and $\bar\g_0$'s, $y=0$ for the Type IIB
superstring, and $y=1$ for the Type IIA superstring.

\section{Relative chiral cohomology in the R-sector}
 
We are guaranteed by the result of [\horowitz] that   
{\it absolute} cohomologies are isomorphic
in all pictures, in particular, this applies to zero-momentum
cohomology in the R sector. 
Moreover, at non-zero momentum, we have shown that
{\it relative} chiral cohomologies are the same in all pictures
(Theorem, section 2). There is no guarantee, however, that at zero momentum
the  {\it relative} cohomologies will be the same. They are not, as we
show now by explicit consideration of the R sector.  

We will calculate the relative cohomology classes for the $-1/2$ and
$-3/2$ pictures. At
zero momentum and with  $L_0 =0$, 
we can only build candidate states using zero modes.
Therefore, the chiral BRST operator in \expbrst\ reduces 
to $Q= -\gamma_0^2 b_0$.
It follows that in the relative complex (where all states are annihilated
by $b_0$) all states are $Q$-closed
and no state can be $Q$-exact.
Thus every (non-zero) 
zero-momentum state annihilated by $b_0$ and $L_0$ represents
a cohomology class. 
The matter
zero modes have been used (the index $\alpha$ on the vacuum), $c_0$ cannot
be used, so we are 
only left with $\beta_0$ and $\gamma_0$.
For the $-1/2$ picture only $\gamma_0$ can be used and 
we therefore find,\foot{Rather than
describing the cohomology classes $H^n(Q,b_0)$ by the corresponding vector
spaces as is customary in the mathematics literature, we simply list
representatives.} 
  
$$\eqalign{H_{-1/2}^{n<0} (Q, b_0) &= 0\,, \cr
H_{-1/2}^{0} (Q, b_0) &= \ket{-\fracs12}^\alpha \,, \cr
H_{-1/2}^{n>0} (Q, b_0) &= \gamma_0^n\ket{-\fracs12}\,. \cr } 
\eqn\rrel$$
Here $H^{n}$ denotes the cohomology class at ghost number $n$. 
There are no classes for negative ghost numbers, one spinor state
at ghost number zero, and one spinor state for every other positive
ghost number. The spinor index in the last equation is $\alpha$ for 
$n$ even, and $\alpha'$ for $n$ odd. On the other hand, for picture 
$-3/2$ we find
$$\eqalign{H_{-3/2}^{n<0} (Q, b_0) &= \b_0^n\ket{-\fracs32}\,, \cr
H_{-3/2}^{0} (Q, b_0) &= \ket{-\fracs32}^{\alpha'} \,, \cr
H_{-3/2}^{n>0} (Q, b_0) &= 0\,. \cr } 
\eqn\rreltwo$$
The spinor index on the right hand side of the first equation
is $\alpha$ for $n$ odd, and $\alpha'$ for $n$ even.
We note that, given the structure of these cohomologies, no picture 
changing operator of definite ghost number could map one cohomology to
the other.
These cohomologies, however, are dual with respect to  
the linear inner product defined by computing the correlation 
funtion of the corresponding operators on a sphere having a
$c_0$ insertion.

\section{Zero-momentum RR semirelative cohomology computations}

In the NS-NS sector of the superstring, 
the only picture where the vacuum satisfies the condition of highest
weight state for the positively moded superghost
oscillators is the $(-1,-1)$ picture.
There is therefore nothing to relate. 
In the NS-R sector there are two available pictures; the $(-1,-3/2)$ picture
and the $(-1,-1/2)$ picture. The zero momentum semirelative cohomologies 
will be proven to be equivalent in section 3.5.   
In the R-R sector, however, there is a surprising difference between the
cohomologies in the four available pictures. In the $(-1/2,-3/2)$ picture (or
$(-3/2,-1/2)$ picture),
there exists
two inequivalent ways of defining cohomology. One of the them is called
`finite cohomology' and will
be proven equivalent to the cohomology in the $(-3/2,-3/2)$ picture.
The other one is called `infinite cohomology' and 
will be proven equivalent to 
the cohomology in the $(-1/2,-1/2)$ picture. 

When computing BRST cohomology for a complex {\cal H}, 
one typically computes ``finite cohomology", in the sense that
the complex is defined to include only vectors that are finite 
linear superpositions of Fock state vectors.  A cohomology class in this
complex must have a representative built as a finite
linear superposition of Fock states, and cannot
 be written as $[Q,\L]$ where $\L$ also
contains a finite number of terms.   Infinite cohomology is defined as
BRST cohomology in a complex where general states include infinite
linear combinations of Fock space states. A cohomology
class in this complex may be represented by a state which is an
infinite superposition of Fock space states, and 
cannot be written as $[Q,\hat \L]$ 
where $\hat \L$ can contain an infinite number
of terms. 

In practice, one computes cohomology by restricting oneself to 
subspaces of states with fixed momentum, fixed ghost number, and 
$L_0$ eigenvalue zero. In the bosonic string theory complex,  
these subspaces are always finite dimensional, and therefore 
finite and infinite cohomologies are
identical as the finite and infinite complexes are the same.
In superstring theory we have the superghosts
zero modes $\beta_0$ and $\gamma_0$, and there exists a double
infinity of objects, $(\beta_0 \overline\gamma_0)^n$, and 
$(\gamma_0 \overline\beta_0)^n$, with $n\geq 0$,
having total ghost number zero and
$L_0$ eigenvalue zero. These objects vanish on the $(-1/2,-1/2)$ and
$(-3/2,-3/2)$ pictures, and thus the relevant subspaces for cohomology
computations in these pictures are always finite dimensional. This
is not the case for the mixed $(-3/2,-1/2)$ and
$(-1/2,-3/2)$ pictures. The objects $(\beta_0 \overline\gamma_0)^n$
do not annihilate the $(-3/2,-1/2)$ vacuum, and similarly the objects 
$(\gamma_0 \overline\beta_0)^n$ do not annihilate the $(-1/2,-3/2)$ 
vacuum. Therefore, in the case of mixed pictures we can distinguish
between finite and infinite cohomology.

Let us now  consider the explicit computation of semirelative cohomologies
at zero momentum in the RR sector of the superstring. No separate 
discussion will be necessary for IIA and IIB superstrings; we will
leave the spinor indices of the vacua unspecified, and they  
can be reconstructed from the GSO condition, as explained at the
end of section 3.1.
At zero momentum, and with $L_0=\overline L_0 =0$ and $b_0^- =0$,
all states are built by acting with 
$\g_0, \bar\g_0, \b_0, \bar \b_0$ and $c_0^+$ on the vacua.    
The BRST operator can then be read from \expbrst, and up to
an irrelevant overall factor reads  
$$Q =  - b_0^+ (\gamma_0^2 + \bar\gamma_0^2 )\, .    \eqn\qbrsta$$
The results of the computations of BRST cohomology are summarized in
the table below ($\gamma_{\pm}$ and $\beta_{\pm}$ are defined 
in (3.11)),
and explanations on the computations follow.

\vskip .1in
$$\hbox{\vbox{\offinterlineskip
\def\strut{\hbox{\vrule height 25pt depth 20pt width 0pt}}
\hrule
\halign{
\strut\vrule#\tabskip 0.1in&
\hfil$#$\hfil &
\vrule#&
\hfil$#$\hfil &
\vrule#&
\hfil$#$\hfil &
\vrule#&
\hfil$#$\hfil &
\vrule#&
\hfil$#$\hfil &
\vrule#\tabskip 0.0in\cr
&  k && H_{-{3\over 2}\, , - {3\over 2}~  }^{(k)} 
&& H_{-{3\over 2}\, , - {1\over 2}~  }^{(k) {fin}}
&& H_{-{3\over 2}\, , - {1\over 2}~  }^{(k)\infty}
&& H_{-{1\over 2}\, , - {1\over 2}~  }^{(k)}
& \cr\noalign{\hrule}
& -n-1 
&& c_0^+ \b_\pm^{n+2}
&& \pm i \g_\mp \b_0^{n+2}    
&& 0\hskip-6pt /  
&& 0\hskip-6pt / & \cr\noalign{\hrule}
&-1 
&&  c_0^+ \b_\pm^2 
&&  \pm i \g_\mp \b_0^2 
&& 0\hskip-6pt /
&&0\hskip-6pt / & \cr\noalign{\hrule}
& 0 
&&   c_0^+ \b_\pm  
&& \pm i \g_\mp \b_0 
&&  {c_0^+\over \bar\gamma_0} \sin (\beta_0 \bar \gamma_0)
&& {\bf 1} &
\cr\noalign{\hrule}
& +1 
&& c_0^+ 
&& \bar\gamma_0 
&& c_0^+ \exp (\pm i\beta_0\bar\gamma_0)
&&  \g_\pm   
&\cr\noalign{\hrule}
& n+1 
&& 0\hskip-6pt /
&& 0\hskip-6pt /
&& c_0^+\bar\gamma_0^n \exp (\pm i\beta_0\bar\gamma_0)
&& \g_\pm^{n+1} 
&\cr\noalign{\hrule} }}}$$
\noindent {\bf Table}. List of semirelative cohomology classes in 
the Ramond-Ramond
sector of the superstring at zero momentum. Shown are the 
standard choices of picture.
\bigskip

\noindent
\underbar{Diagonal Pictures}. 
We begin with  the computation in the $(-1/2,-1/2)$ picture.  
It follows from the form of $Q$ that a state is $Q$-closed
if and only if it involves no $c_0^+$. Since the
$\b_0$ and $\bar \b_0$ modes kill the vacuum in this picture, 
there are no candidate states at negative ghost numbers, 
and as a consequence there is
no cohomology at negative ghost numbers. At ghost number zero,
the vacuum states are all the candidate states, they are
all $Q$-closed, and given the absence of ghost minus one states,
they represent BRST classes. We indicate this as an entry ``one" in the
table, standing for the fact that there are no oscillators acting
on the vacuum. At $G=1$, there are no trivial states, since 
all candidates at $G=0$ were $Q$-closed. Therefore, 
the BRST classes
are represented by $\gamma_0$ acting on the vacuum and $\bar\g_0$
acting on the vacuum. The states where $c_0$ only acts on the vacuum
are unphysical $G=1$ states, so starting at $G=2$ there exist trivial
states. The general pattern at $G\geq 2$ is readily elucidated.
For this purpose  it is 
convenient to introduce new zero modes\foot{The vector 
spaces are naturally
complex, so this change of variables is acceptable.
When writing states in terms of $\pm$ zero modes, the properly
projected GSO states are found by rewriting the states in terms
of the original holomorphic and antiholorphic zero modes, and
applying the rules explained earlier. For example, the bispinor state
$\g_{\pm}|-{1\over 2},-{1\over 2}\>$ in the IIB superstring is 
$\g_0|-\half\>^{\alpha'}\otimes |-\half\>^{\beta}$ 
$\pm i\bar\g_0|-\half \>^\beta\otimes|-\half\>^{\alpha'}$. }   
$$\eqalign{
\g_\pm &= \fracs12 (\g_0 \pm i\bar\g_0)\, ,\cr
  \quad \b_\pm &= \,\,\b_0 \pm i\bar\b_0\, ,\cr  }\eqn\newbass$$
satisfying commutation relations
$$[\gamma_\pm \,,\, \beta_\mp ] = 1 \,,\quad 
[\gamma_\pm \,,\, \beta_\pm ] = 0 \, . \eqn\newcom$$
The BRST operator then reads
$Q \sim b_0^+ \g_+\g_-$. The BRST closed states at $G=n+1$ are represented 
by homogeneous 
polynomials $p_{n+1}(\g_+,  \g_-)$ of degree $n+1$. This is a vector
space of dimension $n+2$. Consider now representatives
for unphysical states at
$G=n$. Any state built by acting on the vacuum with $c_0^+$
and with a monomial $m_{n-1}(\g_+, \g_-)$ of degree $n-1$
is unphysical. Since $Q$ simply deletes the $c_0^+$ and 
multiplies by $\g_+\g_-$, it is clear that on the vector space
spanned by these monomials, the kernel of $Q$ is the zero vector. 
Thus the unphysical states are represented (non-canonically) by
the vector space of  
homogeneous polynomials $p_{n-1}(\g_+, \g_-)$ of degree $n-1$,
a vector space of dimension $n$. It now follows that the
dimension of the physical space at ghost number $(n+1)$ is
given by $(n+2) -n= 2$. Two representatives are readily chosen,
they are $\g_\pm^{n+1}$. It is clear they cannot be trivial as any
trivial state is the sum of monomials each of which contains at least one
$\g_+$ and one $\g_-$. This completes the computation of the 
semirelative cohomology in the $(-1/2,-1/2)$ picture.   

The computations in the $(-3/2,-3/2)$ picture are not all that different.
Candidate states are built acting with $\b_+, \b_-,$ and $c_0^+$. 
We first claim that a state is $Q$-exact if and only if it
has no $c_0^+$. On the one hand, an arbitrary state $\b_+^n \b_-^m \ket{0}$
without $c_0^+$ can be written as
$\sim Q c_0^+\b_+^{n+1} \b_-^{m+1}\ket{0}$. On the other hand, 
any exact state is obtained by $Q$ action, which includes multiplicative
action with $b_0^+$. Since $b_0^+$ annihilates the vacuum, in order to 
get a nonzero result the unphysical state must include a $c_0^+$, and the
trivial state will not include it. All physical states at ghost
number $(-n)$ are therefore of
the form $c_0^+ p_{n+1} (\b_+,\b_-) \ket{0}$, where $p_{n+1} (\b_+,\b_-)$
is an homogeneous polynomial of degree $n+1$. The physical states
are now those annihilated by $\gamma_+\gamma_-$, these are simply the 
states build with either $\b_+$'s or $\b_-$'s but not both. Therefore,
at ghost number $(-n)$ we simply have the states $c_0^+ \b_\pm^{n+1}$.

It is readily seen from the above table that the number of elements in the
$(-3/2,-3/2)$ cohomology at ghost-number $G$ is equal to the number of
elements in the $(-1/2,-1/2)$ cohomology at ghost-number $1-G$. This is
reasonable, as it implies that the
non-degenerate bilinear form 
$$\<\,\Psi_{-3/2,-3/2}\, |\,c_0^- \,| \Phi_{-1/2,-1/2}\>\eqn\norm$$ 
pairing the semirelative complexes ${\cal H}_{-1/2,-1/2}$ and 
${\cal H}_{-3/2,-3/2}$
induces a nondegenerate bilinear form on the zero-momentum
semirelative cohomologies.

\medskip
\noindent
\underbar{Mixed Pictures.} We now focus  
on the $(-3/2,-1/2)$ picture (the results for the $(-1/2,-3/2)$ picture
are obtained simply by exchanging holomorphic and antiholomorphic sectors).
We begin our analysis by calculating cohomologies for $G=n+1\geq 2$, which
requires consideration of $G=n$ states that are not annihilated by
$Q$. These are of the form
$$a_p \equiv c_0^+ {\bar \g}_0^{n-1} (\b_0 \bar\g_0)^p \,\ket{0}\,, 
\quad p=0,1,\cdots \infty\,.
\eqn\firstlist $$
Note that there are an infinite number of states, due to the fact
that $(\b_0 \bar \g_0)$ does not kill the vacuum, is Grassmann even,
has ghost number zero, and $L_0= \bar L_0=0$. Since
$\g_0$ kills the vacuum and $[\g_0,\b_0]=1$, we can write
the BRST operator
as 
$$  Q \sim  b_0^+ \Bigl( {\partial^2\over \partial\b_0^2} + \bar\g_0^2 \Bigr)
\, . \eqn\brft$$
It is clear that $Qa_p \not= 0$ for any fixed $p$, and given that $Q$
includes multiplication by $\bar\g_0^2$, $Q$ cannot annihilate any finite
linear combination of $a_p$'s.  
The candidate states at $G=n+1\geq 2$ are of the form
$$b_p \equiv {\bar \g}_0^{n+1} (\b_0 \bar\g_0)^p \,
\ket{-\fracs32,-\fracs12}\,, \quad
c_p \equiv c_0^+ \,{\bar \g}_0^{n}\, (\b_0 \bar\g_0)^p \,
\ket{-\fracs32,-\fracs12}\,, 
\quad p=0,1,\cdots \infty\,.
\eqn\seclist $$
It follows from \brft\ and \firstlist\ that
$$\eqalign{
Qa_0 &= b_0\,, \cr
Qa_1 &= b_1\,, \cr
Qa_p &= b_p + p(p-1) b_{p-2} \,,\quad p \geq 2\,, \cr} \eqn\comhg$$
showing that all $b_p$ are trivial as they can all be written as
$Q$ acting on a finite linear combination of $a_{p'}$'s with
$p'\leq p$. On the other hand none of the $c_p$'s can be exact,
since exact states cannot have a $c_0^+$. As remarked above, no finite
linear combination of $c_p$'s can be $Q$ closed, but it is clear that
a state of the form  $c_0^+ \bar\g_0^n \, h( \b_0\bar\g_0)$, 
where $h(x)$ is simply
a function of a single variable, will be $Q$-closed if $h''+ h =0$. 
We therefore
find two solutions
$$  c_0^+ \bar\g_0^n \, \sin ( \b_0\bar\g_0)\,,\, \hbox{and}\quad 
 c_0^+ \bar\g_0^n \, \cos ( \b_0\bar\g_0)\,\quad n\geq 1\,. \eqn\infcohr$$ 
In summary, for $G=n+1\geq 2$ we have 
found no finite cohomology and two states in the infinite
cohomology.

The case of $G=1$ is slightly different with regards to the finite
cohomology. The $G=0$ states (in the finite complex) that are not annihilated
by $Q$ are of the form
$$d_p \equiv c_0^+ \b_0\,  (\b_0 \bar\g_0)^p \,\ket{-\fracs32,-\fracs12}\,, 
\quad p=0,1,\cdots \infty\,,
\eqn\firrstlist $$
and the states at $G=1$ 
that could be in finite cohomology
are 
$$e_p \equiv {\bar \g}_0 \, (\b_0 \bar\g_0)^p \,\ket{-\fracs32,-\fracs12}\,,  
\quad p=0,1,\cdots \infty\,.
\eqn\secclist$$
It it then simple to find 
$$\eqalign{
Qd_0 &= e_1\,, \cr
Qd_1 &= e_2 + 2e_0\,, \cr
Qd_p &= e_{p+1} + p(p+1) e_{p-1} \,,\quad p \geq 2\,. \cr} \eqn\xcomhg$$
It is clear that all $e_k$ with $k$ odd are trivial. On the other hand
$e_0 = \bar\g_0\ket{-\fracs32,-\fracs12}$ is not trivial in the 
finite complex. It is 
trivial on the infinite complex 
since, using the above equations for $p$ 
odd, we can write $e_0 = \fracs12 Q (d_1 - \fracs1{12} d_3 + \cdots)$.
This is an example of a phenomenon we find here; every cohomology
class in the finite complex is trivial on the infinite complex.

The computations for other ghost numbers follow the above
lines quite closely so they will not be discussed explicitly.
The results are indicated on the table. We mention that for 
$G= -n-1 $, we find two states 
$$ \b_0^{n+1}\ket{-\fracs32, -\fracs12}\,, \quad\hbox{and}\quad
\b_0^{n+2}\bar\g_0\ket{-\fracs32, -\fracs12}\,, \eqn\hjk$$ 
but for later purposes it is convenient to express them as the
linear combinations
$$[\pm i (n+2) \b_0^{n+1} + \b_0^{n+2} \bar\g_0 ] \ket{-\fracs32, -\fracs12}
= \pm \, i\, \g_\mp\,  \b_0^{n+2}\ket{-\fracs32, -\fracs12}\,, \eqn\inta$$
as indicated on the table. Note that the pattern actually holds
for $G=1$ (corresponding to $n=-2$)
since $\pm i \g_\mp\ket{-\fracs32, -\fracs12} = 
\bar\g_0 \ket{-\fracs32, -\fracs12}$. 

\bigskip
\noindent
\underbar{Comments.} In our conventions the string  
field has ghost number zero, thus
a D-brane state, which is naturally a bra that couples to the
string field, can be thought as a ket of ghost number plus one.
If the string field is represented in the $(P,\bar P)$ picture,
the D-brane state must be in the $(-2-P, -2-\bar P)$ picture.
The superghost dependence of D-brane states has been discussed
in [\yost] where it was shown that 
$$\ket{B, \pm } = \exp( \mp i \b_0\bar\g_0) \,
\exp \Bigl( \pm i \sum_{r\geq 1} (\g_{-r}\bar \b_{-r} - \b_{-r}\bar \g_{-r})
\Bigr)\ket{-\fracs32,-\fracs12 }\,. \eqn\dbrane$$
These states are $Q$-closed, and we note that the $L_0=\bar L_0=0$ part
are exactly the infinite cohomology classes we have found (and the rest
of the states is $Q$-exact). They are appropriate 
electric and magnetic D-branes for a string field in the 
$(-1/2, -3/2)$  
picture. 
The isomorphisms of cohomology stated in \relation,  to be
established in the next section, imply that
$G=1$ cohomology
in the $(-1/2,-1/2)$ picture describes the same cohomology
classes that are associated to the D-brane states.
\foot{This is different from the observation
of [\yost] that the above D-brane states
can be rewritten in the form $\ket{B, \pm}= 
c_0^+ \delta(\gamma_\pm)\ket{-\half,-\half}$.
This rewriting does not imply that the states $\ket{B, \pm}$
belong to the $(-1/2,-1/2)$ picture, since delta functions of
superghosts are indeed picture changing operators. }

Similarly, electric and magnetic
gauge fields at zero-momentum are represented by $G=0$ elements of 
the finite $(-3/2,-1/2)$ cohomology
(or $(-3/2,-3/2)$ cohomology, via
\relation).  This is consistent
with the non-degeneracy of \norm\ since 
zero-momentum electric and magnetic gauge fields have
non-zero coupling to electric and magnetic D-branes [\sagn].    
Note  that the $G=0$ elements of $(-1/2,-1/2)$ cohomology
(or $(-3/2,-1/2)$ infinite cohomology) 
are field-strengths rather than gauge fields since they
are invariant under $x$-dependent
gauge-transformations. 

\section{Isomorphisms of semirelative cohomology}

The explicit results of the previous section, as summarized
in the table, suggest natural isomorphisms of the zero momentum
semi-relative cohomologies at the various pictures. Since the 
relevant vector spaces defined by the cohomology classes
are finite dimensional, naive counting shows that the vector
spaces obey the isomorphism indicated in \relation. It is the
purpose of the present section to construct a picture-raising 
operator and a
picture-lowering operator 
that induce a ``canonical" isomorphism 
between the various cohomologies. This means that 
the matching of physical states
it is not just a counting coincidence, but that the states
related by the canonical isomorphism  are really physically
equivalent.

The picture-raising operator $X_0$ is defined as in \threex, and we can 
can read out the zero mode piece by acting on zero-momentum
states in the ${-{3\over 2}}$ picture. 
One can readily show that $X_0 c_0\ket{-\fracs32}^\alpha 
 =\g_0 \ket{-\fracs12}^\alpha$ , 
$X_0 c_0\beta_0\ket{-\fracs32}^\alpha  =  \ket{-\fracs12}^\alpha $, 
and $X_0$ annihilates all other states. 
Making use of the familiar relation $\delta(\beta_0) \ket{-\fracs32}^\alpha 
= \ket{-\fracs12}^\alpha $, we find that 
$$X_0 c_0\ket{-\fracs32}^\alpha = b_0 \, \{\g_0 , \delta (\b_0) \} 
c_0\ket{-\fracs32}^\alpha ,
\quad
X_0 c_0\beta_0\ket{-\fracs32}^\alpha  
= b_0 \, \{\g_0 , \delta (\b_0) \} c_0\beta_0\ket{-\fracs32}^\alpha
\,,\eqn\ritz$$
where $\{ \cdot , \cdot \}$ denotes the anticommutator. 
Note that  
$\delta(\beta_0) \ket{-\fracs32}^{\alpha'}  
= - \ket{-\fracs12}^{\alpha'} $ since $\delta(\beta_0)$ is fermionic
and therefore anti-commutes with $\psi^\mu_0$.   

{} From \ritz, we are led to consider the 
picture-raising 
operator  
$${\cal X}_0 \equiv b_0^+ \, \{\g_0 , \delta (\b_0) \}\,, \eqn\zmpr$$
which is built exclusively of zero modes. In the zero-momentum
$L_0=0$ sector, $Q\sim b_0\g_0^2+\bar b_0 \bar\g_0^2$,   
and therefore 
${\cal X}_0 Q = Q {\cal X}_0 = 0$. Also, ${\cal X}_0$ clearly commutes
with $b_0^-$ and therefore defines a map of semirelative
closed string cohomologies. 

For the picture-lowering operator, consider the usual operator 
$Y= c\partial \xi e^{-2\phi}$, and its zero mode piece
$Y_0 = \oint {dz\over 2\pi iz} Y(z)$. The operator $Y_0$
involves both zero modes and non-zero modes, and we can read
out the zero mode piece by acting on zero-momentum
states in the ${-\half}$ picture. 
One can readily show that $Y_0 \ket{-\half}^\alpha  = 
c_0\b_0 \ket{-\fracs32}^\alpha $, 
$Y_0 \gamma_0\ket{-\half}^\alpha  = c_0\ket{-\fracs32}^\alpha $,
and $Y_0$ annihilates
all other states.  
Making use of the relation $\delta(\gamma_0) \ket{-\half}^\alpha 
= \ket{-\fracs32}^\alpha $ (which implies that
$\delta(\gamma_0) \ket{-\half}^{\alpha'}   
= - \ket{-\fracs32}^{\alpha'} $), we find that 
$$Y_0 \ket{-\fracs12}^\alpha = c_0 \, [\b_0 , \delta (\g_0) ] 
\ket{-\fracs12}^\alpha ,
\quad
Y_0 \gamma_0\ket{-\fracs12}^\alpha  
= c_0 \, [\b_0 , \delta (\g_0) ] \gamma_0\ket{-\fracs12}^\alpha .  \eqn\bnv$$
{}From this, we are led to consider the picture-lowering 
operator   
$${\cal Y}_0 \equiv c_0 \, [\b_0 , \delta (\g_0) ]\,, \eqn\zmpc$$
which is built exclusively of zero modes. One readily verifies that
${\cal Y}_0 Q = Q {\cal Y}_0 = 0$. Indeed, since
$Q\sim b_0\, \g_0^2$, and given that ${\cal Y}_0$ has a single 
$\b_0$, the
$\gamma_0^2$ factor of $Q$ can be made to hit the delta 
function $\delta(\g_0)$. The operator ${\cal Y}_0$, however, does
not commute with $b_0$. This had to be the case, for otherwise
we could prove an isormorphism of the zero-momentum chiral relative
cohomologies, in direct contradiction with the explicit computation
discussed in section 3.2.

For the closed string case, 
we can attempt to set
$${\cal Y}_0 \sim  c_0^+
 \, [\b_0 , \delta (\g_0) ]  \, \,  \eqn\zmpcc$$ 
and while this operator commutes with $b_0^-$, it does not anymore
commute with the complete BRST operator which 
reads $Q\sim b_0^+ (\g_0^2 + \bar\g_0^2)$. 
In order to obtain an
operator that commutes with $Q$, 
we begin by introducing some notation.
Let
$$B^1\equiv [\b_0, \delta (\g_0)]\, , \quad  B^2\equiv [\b_0, B^1]
\,,\cdots\,\,
\quad B^{n+1} = [\b_0 , B^n\,]\,. \eqn\recb$$ 
It is straightforward to verify that
$$[B^n , \g_0 ] = 0\,, \quad [B^n , \bar\g_0 ] = 0\,, \eqn\tryout$$
where the second equation is trivially satisfied. It is also simple
to verify by induction that
$$\g_0 \,B^n = n\, B^{n-1} \, , \eqn\canceld$$
which indicates that formally, thinking of the superscript of $B$ as 
an exponent, $\g_0 \sim \partial/\partial B$. Finally, acting on the
vacua we have
$$B^n\ket{-\fracs12} =\pm\, \b_0^n \ket{-\fracs32}\,,  \eqn\baction$$  
where the top sign applies to an $\alpha$ spinor index and 
the bottom sign applies to an $\alpha'$ spinor index.

We now claim that 
$${\cal Y}_0 \equiv  c_0^+ \cdot {1\over \bar\g_0} \cdot 
\sin ( \bar\g_0 \,B) \,, \eqn\itis$$
is the desired picture-lowering operator which commutes 
with $Q$.
In the power
series expansion of sin we take $(B)^n \equiv B^n$, as suggested above.
The first term in the series expansion is the operator indicated in \zmpcc\
and the other terms are the corrections necessary to
make the operator commute with $Q$. Indeed, $[{\cal Y}_0, b_0^-] =0$
is manifest, and
$[Q, {\cal Y}_0] =0$ simply requires  
$$(\g_0^2 + \bar \g_0^2 ) \sin ( \bar\g_0 \,B) = 0\,, \eqn\reqint$$
where we made use of relations \tryout.  But this equation is manifestly
satisfied by virtue of the identification of $\g_0$ with $\partial/\partial B$. 
This confirms that the inverse picture changing
operator ${\cal Y}_0$ in \itis\ defines a 
map of semirelative cohomologies at zero momentum. Is is also clear
from the explicit expressions that the operators ${\cal Y}_0$
and ${\cal X}_0$ commute with both left and right GSO projections.

In order to prove isomorphisms of cohomology  we will 
verify explicitly that ${\cal Y}_0$ defines a one to one
surjective map of cohomologies. In fact, we will also see
that ${\cal X}_0$ defines an explicit operator inverse for ${\cal Y}_0$.
Let us first show that
$${\cal Y}_0 :  H^{(k)}_{-\fracs12,-\fracs12} \to H^{(k)\infty}_{-\fracs32,
-\fracs12} \eqn\fmapc$$ 
is a one to one surjective map. By construction, 
\itis\ does indeed
map correctly the cohomology at $G=0$ (see table). At $G=1$ we have
$$\eqalign{
{\cal Y}_0 \,\,\g_\pm \ket{-\fracs12,-\fracs12} &= 
{\cal Y}_0\, (\g_0 \pm i \bar \g_0 ) \ket{-\fracs12,-\fracs12} \cr 
&=(\g_0 \pm i \bar \g_0 ){\cal Y}_0\ket{-\fracs12,-\fracs12}  \cr
&= \pm \,c_0^+ {1\over \bar\g_0} 
(\g_0 \pm i \bar \g_0 )\sin (\b_0\bar\g_0) \ket{-\fracs32,-\fracs12}  \cr
&= \pm \,c_0^+ \exp (\pm i\b_0\bar\g_0) \ket{-\fracs32,-\fracs12}\,,  \cr } 
\eqn\mapch$$
where use was made of \tryout\ and \baction, and the $\pm$ 
must be chosen depending on the spinor type of the holomorphic  
vacuum in the  ket appearing in the left hand side.\foot{Note    
that this equation, as written, requires a specific ket and therefore does
not incorporate the GSO condition. This is not a problem since the proper
$\g_\pm \ket{-\half,-\half}$ states in any theory are the GSO projection of
$\g_\pm\, [\,\, \ket{-\half,-\half}^{\alpha\beta}+
 \ket{-\half,-\half}^{\alpha'\beta}
+ \ket{-\half,-\half}^{\alpha\beta'}
+ \ket{-\half,-\half}^{\alpha'\beta'}]$. The action of ${\cal Y}_0$
on this state follows immediately from \mapch, and GSO projection
can be applied then.} In the last step,
we recognized
that $\g_0$ acts as $\partial/\partial \b_0$. Comparing with the table,
we see that ${\cal Y}_0$ does act correctly at $G=2$. For higher ghost
numbers the verification is now very simple 
$$\eqalign{
{\cal Y}_0 \,\,\g_\pm^{n+1} \ket{-\fracs12,-\fracs12} &= 
\g_\pm^{n}\, {\cal Y}_0 \,\,\g_\pm \ket{-\fracs12,-\fracs12}\cr
&= \pm\, c_0^+  (\g_0 \pm i \bar \g_0 )^n  \exp (\pm i\b_0\bar\g_0)
\ket{-\fracs32,-\fracs12} \,, \cr
&\sim\,  \pm \, c_0^+  \bar \g_0^n  \exp (\pm i\b_0\bar\g_0)
\ket{-\fracs32,-\fracs12} \,, \cr } 
\eqn\mapchd$$
which is the expected result. This confirms our claims about \fmapc.

Let us now show that the map  
$${\cal X}_0 :  H^{(k)\infty}_{-\fracs32,-\fracs12} \to H^{(k)}_{-\fracs12,
-\fracs12} \eqn\gmapc$$ 
provides an inverse for \fmapc. At $G=0$, only the first term 
in the Taylor expansion of  
${{c_0^+}\over {\bar\g_0}} \sin(\b_0\bar\g_0)\ket{-\fracs32,-\fracs12} $ 
survives when hit by 
${\cal X}_0$, and it is easy to check that the image of this term is   
$\pm\ket{-\fracs12,-\fracs12} $, where once more, the $\pm$ choice
depends on the
spinor type of the holomorphic vacuum. 
At $G=1$, the first two terms in the
Taylor expansion of the state $c_0^+\exp (\pm i\b_0\bar\g_0)   
\ket{-\fracs32,-\fracs12}$ 
survive when hit by  
${\cal X}_0$, and these two terms combine to give 
$\pm \g_{\pm}\ket{-\fracs12,-\fracs12} $. 
At $G>1$, the first two terms in
the Taylor expansion of the states
survive to give $\pm \,\bar\g_0^n (\g_0 \pm i \bar\g_0)  
\ket{-\fracs12,-\fracs12} $, which are in the same cohomology class as 
$\pm \bar\g_\pm^{n+1} \ket{-\fracs12,-\fracs12} $.  
This confirms our claims about \gmapc.

We now want to verify that  
$$\overline{\cal Y}_0 :  H^{(k)\, fin}_{-\fracs32,-\fracs12} 
\,\to \, H^{(k)}_{-\fracs32,
-\fracs32} \eqn\fmapcc$$
is a one to one surjective map. Here, as we must change the picture
of the right-movers, we consider the operator 
$$\overline {\cal Y}_0 \equiv  c_0^+ \cdot {1\over \g_0} \cdot 
\sin ( \g_0 \,\bar B) \,, \eqn\itiss$$
which also commutes with $b_0^-$ and with the BRST operator.
The first nontrivial check is at $G=1$ where we find 
$$\eqalign{
\overline{\cal Y}_0  \bar\g_0 \ket{-\fracs32, -\fracs12} 
&= c_0^+ \,\bar B \,\bar\g_0 \ket{-\fracs32, -\fracs12} \cr
&= c_0^+ \,\bar\g_0 \, \bar B \ket{-\fracs32, -\fracs12} \cr
&= \pm\, c_0^+ \,\bar\g_0 \, \bar \b_0 \ket{-\fracs32, -\fracs32} \cr
&= \pm\, c_0^+  \ket{-\fracs32, -\fracs32}\,, \cr} \eqn\bvghx$$ 
in agreement with the result of the table. Note that the power
expansion of $\overline{\cal Y}_0 $ collapsed to the first term
since $\g_0$ annihilates the vacuum, and there are no $\b_0$'s in the
state upon consideration. We now consider the general term at
$G= -n-1$, this time we must evaluate
$$ c_0^+ {1\over \g_0} 
\sin ( \g_0 \,\bar B)\,\, [\,\pm i \,(n+2)\, \b_0^{n+1} +
 \b_0^{n+2}\, \bar\g_0\, ]\, \ket{-\fracs32, -\fracs12} \eqn\meval$$ 
Consider now the case when $n=2l$ is even (the odd case can be treated
similarly). Using the series expansion for sin we find
$$\pm \, c_0^+  \sum_{k=0}^\infty {(\pm i)^{2k}\over (2k+1)!} \, \g_0^{2k}
[\,\pm i \,(2l+2)\, \b_0^{2l+1} +  
 \b_0^{2l+2}\, \bar\g_0\, ]\,\bar\b_0^{2k+1} \ket{-\fracs32, -\fracs32} $$
where we used $[\bar B^n, \bar \g_0]=0$. Here both $\g_0$ and $\bar \g_0$,
since they kill the vacuum,
act as derivatives and the infinite sum is truncated to
$$\pm \, c_0^+ \,\Bigl[\,\, \sum_{k=0}^l \Bigl( {2l+2\atop 2k+1} 
\Bigr)\b_0^{2l+1-2k} (\pm i\bar\b_0)^{2k+1}
+  \sum_{k=0}^{l+1} \,\Bigl( {2l+2\atop 2k} \Bigr)
 \b_0^{2l+2-2k} \, (\pm i\bar\b_0)^{2k} \,\,\Bigr] \ket{-\fracs32, -\fracs32} $$
The two sums can now be combined into a single sum
$$\pm \, c_0^+ \, \sum_{k'=0}^{2l+2} \Bigl( {2l+2\atop k'}
\Bigr)\b_0^{2l+2-k'} (\pm i\bar\b_0)^{k'}
 \ket{-\fracs32, -\fracs32} 
=\pm\, c_0^+(\b_0\pm i\bar\b_0)^{2l+2}\ket{-\fracs32, -\fracs32}\,,\eqn\vb$$
and this is recognized as the state of $G=-n-1$ in the $(-3/2, -3/2)$
semirelative cohomology. This verifies that \fmapcc\ induces the
claimed isomorphism.

Similarly, it is straightforward to verify that 
$$\overline{\cal X}_0 :  H^{(k)}_{-\fracs32,-\fracs32} 
\,\to \, H^{(k)\, fin}_{-\fracs32,
-\fracs12} \eqn\fmapcc$$
acts as the inverse of $\overline{\cal Y}_0$. 

\section{Semirelative classes for Heterotic R-sector, and R-NS superstrings}

In this section, 
we consider first the R-sector of heterotic strings. While
the left movers here define a bosonic string, the right movers can be
in either the $-1/2$ or $-3/2$ pictures. We will compute zero momentum
cohomologies in both pictures and show that the cohomologies are equivalent
by constructing a picture changing operator that gives an explicit 
isomorphism.

The antiholomorphic sector here is
 that of  bosonic strings and will be considered only at zero momentum,
and with $\bar L_0 = 0$. The vacuum $\ket{\Omega} = \bar c_1 \ket{{\bf 1}}$,
has $\bar L_0 = -1$ and is annihilated by all positively moded oscillators.
The constraint $\bar L_0 =0$ implies that states must have an oscillator
of mode number $-1$ acting on $\ket{\Omega}$, and oscillators with mode number
less than $-1$ cannot be used. Under those circumstances, 
the relevant 
antiholomorphic part of BRST
operator reads $\bar Q \sim \bar b_0 \bar c_{-1} \bar c_{1}$. Now consider
using the $(-1/2)$ picture for the holomorphic sector. The complete vacuum
will be denoted as $\ket{-\fracs12}_H \equiv \ket{-\fracs12}\otimes\ket{\Omega}$
where the $H$ stands for heterotic. Its ghost number is defined to be zero.
The semirelative
complex here is built by action with the set of 
oscillators $\{\g_0, c_0^+, \bar b_{-1}, \bar c_{-1},
\bar \alpha_{-1}^\mu \}$ on the vacuum. The BRST operator, using a 
specific (but conventional)
relative normalization between left and right sectors reads
$$Q= b_0^+ ( \bar c_{-1} \bar c_{1} + \g_0^2 ) \,. \eqn\hetbrst$$
A straightforward computation gives that the only nonvanishing 
cohomology classes are 
$$\eqalign{H^{-1} (Q, b_0^-) &= \bar b_{-1} \ket{-\fracs12}_H \,, \cr
H^{~0}\, (Q, b_0^-) &= \bar \alpha_{-1}^\mu \ket{-\fracs12}_H \,, \quad
\bar b_{-1} \g_0 \ket{-\fracs12}_H  \,, \cr
H^{+1}\, (Q, b_0^-) &= \bar\alpha_{-1}^\mu\,\g_0 \ket{-\fracs12}_H\,,
\quad (\bar c_{-1} - \bar b_{-1} \g_0^2) \ket{-\fracs12}_H\,, \cr
H^{+2} (Q, b_0^-) &= \gamma_0 \,(\bar c_{-1} - \g_0^2 \,\bar b_{-1})
\ket{-\fracs12}_H\,. \cr } 
\eqn\rrelhet$$
The computations for the (-3/2) picture uses the complex built on
$\ket{-\fracs32}_H \equiv \ket{-\fracs32}\otimes\ket{\Omega}$ by 
the action of the same set of oscillators, except that $\g_0$ is 
replaced by $\b_0$. The answer this time is
$$\eqalign{H^{-1} (Q, b_0^-) &= c_0^+ (\,\bar b_{-1}\b_0
-\fracs16 \bar c_{-1} \b_0^3)  \ket{-\fracs32}_H \,, \cr
H^{~0}\, (Q, b_0^-) &= c_0^+\bar \alpha_{-1}^\mu \b_0\, 
\ket{-\fracs32}_H \,, \quad
c_0^+ (\, \bar b_{-1}-\fracs12 \bar c_{-1} \b_0^2) \ket{-\fracs32}_H  \,, \cr
H^{+1}\, (Q, b_0^-) &= c_0^+\bar\alpha_{-1}^\mu\, \ket{-\fracs32}_H\,,
\quad c_0^+\,\bar c_{-1}\b_0 \ket{-\fracs32}_H\,, \cr
H^{+2} (Q, b_0^-) &= c_0^+ \,\bar c_{-1}\,
\ket{-\fracs32}_H\,. \cr } 
\eqn\rrelhetpu$$
Comparing the last two lists, 
we see that the dimensionalities of the
cohomologies agree. More importantly, using the methods of the previous
subsection we can construct  picture-raising and picture-lowering
operators ${\cal X}_0^H$ and 
${\cal Y}_0^H$. 
The operator ${\cal X}_0^H$ is defined as in \zmpr\ while 
$${\cal Y}_0^H = c_0^+ \Bigl( B - \fracs16 \bar c_{-1} \bar c_1 B^3\Bigr) \,,
\eqn\getpic$$
where this operator manifestly commutes with $b_0^-$ and a short computation
shows that it commutes with the BRST operator indicated in \hetbrst.
It is straightforward to show that ${\cal Y}_0^H$ acting on the list
\rrelhet\ gives us precisely the list in \rrelhetpu. This confirms that
the two pictures of the heterotic string contain the same physical
zero momentum states. Since that is also the case for non-zero momentum,
there is no ambiguity in the R sector of the heterotic string.

Consider now the R-NS sector of closed superstrings. 
The antiholomorphic sector here is
NS and will be considered only at zero momentum,
and with $\bar L_0 = 0$. The vacuum $\ket{\Omega}_{NS} = \bar c_1 \ket{-1}$,
is based on the $-1$ picture, has $\bar L_0 = -1/2$, and is annihilated 
by all positively moded oscillators.
The constraint $\bar L_0 =0$ implies that states must have an oscillator
of mode number $-1/2$ acting on $\ket{\Omega}_{NS}$, and oscillators 
with mode number
less than $-1/2$ cannot be used. Under those circumstances the relevant 
antiholomorphic part of BRST
operator reads $\bar Q \sim \bar b_0 \bar \g_{-1/2}\, \bar \g_{1/2}$. 
Now consider
using the $(-1/2)$ picture for the holomorphic sector. The complete vacuum
will be denoted as $\ket{-\fracs12, -1}\equiv \ket{-\fracs12}
\otimes\ket{\Omega}_{NS}$. Its ghost number is defined to be zero.
The semirelative
complex here is built by action with the set of 
oscillators $\{ \g_0, c_0^+, \bar \psi^\mu_{-1/2}, \bar \b_{-1/2},
\bar \g_{-1/2} \}$ on the vacuum. The BRST operator, using a 
specific (but conventional)
relative normalization between left and right sectors reads
$$Q= 
b_0^+ ( \bar \g_{-{1\over 2}} \bar \g_{{1\over 2}} + \g_0^2 ) \,.\eqn\nsrbrst$$
A straightforward computation gives that the only nonvanishing 
cohomology classes are 
$$\eqalign{H^{-1} (Q, b_0^-) &= \bar \b_{-\half} \ket{-\fracs12, -1} \,, \cr
H^{~0}\, (Q, b_0^-) &= \bar\psi_{-\half}^\mu \ket{-\fracs12, -1} \,, \quad
\bar \b_{-\half} \g_0 \ket{-\fracs12, -1}  \,, \cr
H^{+1}\, (Q, b_0^-) &= \bar\psi_{-\half}^\mu\,\g_0 \ket{-\fracs12, -1}\,,
\quad (\bar \g_{-\half} - \bar \b_{-\half} \g_0^2) \ket{-\fracs12, -1}\,, \cr
H^{+2} (Q, b_0^-) &= \gamma_0 \,(\bar \g_{-\half} - \g_0^2 \,\bar \b_{-\half})
\ket{-\fracs12, -1}\,. \cr } 
\eqn\rrelnsr$$
The computations for the (-3/2) R-picture uses the complex built on
$\ket{-\fracs32, -1} \equiv \ket{-\fracs32}\otimes\ket{\Omega}_{NS}$ by 
the action of the same set of oscillators, except that $\g_0$ is 
replaced by $\b_0$. The answer this time is
$$\eqalign{H^{-1} (Q, b_0^-) &= c_0^+ (\,\bar \b_{-\half}\b_0
-\fracs16 \,\bar \g_{-\half}\, \b_0^3)  \ket{-\fracs32, -1} \,, \cr
H^{~0}\, 
(Q, b_0^-) &= c_0^+\bar \psi_{-\half}^\mu \,\b_0\, \ket{-\fracs32, -1} \,,
 \quad
c_0^+ (\, \bar \b_{-\half}-\fracs12\, \bar \g_{-\half}\, \b_0^2)
\ket{-\fracs32, -1} \,, \cr
H^{+1}\, (Q, b_0^-) &= c_0^+\,\bar\psi_{-\half}^\mu\, \ket{-\fracs32, -1}\,,
\quad c_0^+\,\bar \g_{-\half}\,\b_0 \ket{-\fracs32, -1}\,, \cr
H^{+2} (Q, b_0^-) &= c_0^+ \,\bar \g_{-\half}\,
\ket{-\fracs32, -1}\,. \cr } 
\eqn\rrelnsru$$
Once again the last two lists have the same number of cohomology classes
at each ghost number. The picture-raising operator is the same as before
while the picture-lowering operator reads  
$${\cal Y}_0^{NS} = c_0^+ \Bigl( B - \fracs16 \bar \g_{-\half} \bar \g_{\half}
 B^3\Bigr) \,,
\eqn\getpicnsr$$
where this operator manifestly commutes with $b_0^-$ and a short computation
shows that it commutes with the BRST operator indicated in \nsrbrst. This
last computation requires noting that $\bar\g_{\half}^2$ annihilates all
relevant states ({\it i.e.}, $\bar L_0=0$, zero-momentum states).
It is straightforward to show that ${\cal Y}_0^{NS}$ acting on the list
\rrelnsr\ gives us precisely the list in \rrelnsru. This confirms that
the two pictures of the R-NS sector of the  
superstring contain the same physical
zero momentum states. Since that is also the case for non-zero momentum,
there is no ambiguity in the R-NS sector of the superstrings.
All in all, this section showed that inequivalent
cohomologies only occur in the RR sector of superstrings. 

\chapter{Zero momentum $x_0$-cohomology}

In principle,
 a computation of $x_0$-cohomology requires a complete reconsideration
of the earlier computations. Both  
the Fock spaces of fields and gauge parameters must
be extended 
to include multiplication 
by arbitrary  finite order polynomials on the zero modes
$x^\mu_0$ of the non-compact bosonic coordinates.  As usual, comparing with the
previous calculation where $x_0$ was not included, we can both lose physical
states,
due to new gauge parameters, or gain states, due to new fields.
Given the results of  [\astashkevich], 
we
expect 
that the physically relevant  $x_0$-cohomology at zero momentum is a subset 
of the zero-momentum cohomology in the original semirelative complex. 
There is one 
predictable exception to this;  at ghost number minus one,  $x_0$-cohomology
includes new states
linear in $x_0$, these give rise to Lorentz symmetries. 
The present analysis will only consider the zero momentum physical
states found in the previous section and we will ask which ones can be gauged 
away in $x_0$-cohomology.  The states that cannot be gauged away are definitely
$x_0$-cohomology classes, but there could be additional classes represented by
states that contain factors of $x_0$. 

For the purposes of the
present  section, the  relevant BRST operator reads
$$Q = c_0^+ p^2  - b_0^+ (\gamma_0^2 + \bar\gamma_0^2 )
+ (\gamma_0 \,\psi_0^\mu 
+ \bar\gamma_0 \, \bar\psi_0^\mu ) \, p_\mu\, ,    \eqn\qbrsttwo $$
where the zero modes $\psi_0^\mu$ act on the vacua as indicated in
\zmactiona. This BRST operator can be related to that of \qbrst\
by rescaling $c\to \half c$, $b \to 2b$ ,   
$\gamma\to \half \gamma$ and $\beta \to 2 \beta.$

Since the picture changing operators discussed in the previous section
commute with the above BRST operator $Q$,
the isomorphisms of cohomology 
will hold for $x_0$-cohomology; the  $(-{3/2},-{3/ 2}\,)$  
$x_0$-cohomology
will coincide with the finite $(-{3/ 2},-{1/ 2})$  
$x_0$-cohomology, and the infinite
$(-{3/2},-{1/2})$  $x_0$- cohomology will coincide with the 
$(-{1/ 2},-{1/ 2})$ $x_0$-cohomology.
We will therefore only compute the  finite $(-{3/ 2},-{1/2})$ 
cohomology 
and the $(-{1/2},-{1/ 2})$  cohomology; in both cases considering 
Types I, IIA, and IIB superstrings. Our results are summarized
in the table at the end of this section.

\section{Anomalies, Fischler-Susskind and $x_0$-cohomology}

Here we wish to make some comments that relate to the appearance
of anomalies in string theory. These remarks give the appropriate
BRST cohomological framework for the observations first made in
Ref.[\polchinskicai].  
We claim that a string theory
has candidate anomalies if the $x_0$-cohomology at ghost number
plus one is non-vanishing. Here, as in the rest of this paper,
we are using the convention that the physical string field is at
ghost number zero. In a nutshell, the anomaly arises because we
are required to solve an equation of the type $Q\ket{\Phi_0} = -\ket{D}$,
where $\ket{\Phi_0}$ is of ghost number zero. Consistency of this equation
implies that $\ket{D}$ is exact. If the appropriate cohomology at
ghost number one vanishes, $\ket{D}$ is necessarily exact and thus the
above equation has a solution. But if there is nontrivial cohomology 
at $G=1$, there is a potential anomaly. We have an anomaly
if $\ket{D}$ is 
a representative of a nontrivial class since then
the above equation does not have a solution. 
Let us first explain how the equation
arises, and second, why one should use $x_0$-cohomology.

A consistent quantum background in string field theory is 
a background where the effective string action 
 has vanishing one point functions. 
As usual, the effective action is obtained by computing 
(using the Batalin-Vilkovisky
quantum master action) all one-particle (string!)
irreducible graphs. This effective action takes the form
$$\Gamma (\Psi) \sim \fracs12
 \langle \Psi , Q \Psi \rangle  +  \langle\Psi, D\rangle
+ \cdots \,, \eqn\effaction$$
where we have shown the kinetic term, and a possible one-point interaction
that could arise in the computation of the effective action. In order
to have a consistent background, 
the  linear term must vanish. To achieve
this, one must shift the background,
the essence of the Fischler-Susskind mechanism [\fischlersusskind], 
and one tries $\Psi\to \Psi + \Psi_0$, where $\Psi_0$ represents
a set of vacuum expectation values to be determined. 
It follows immediately from the
above equation that the condition that the linear term vanish after the
shift is simply $Q\ket{\Psi_0} = -\ket{D}$. Note that in a momentum 
conserving theory, $\ket{D}$ must be concentrated at zero-momentum. 

The choice of $x_0$-cohomology is the obvious choice, as this is the
choice that allows giving vacuum
expectation values that are polynomials in the spacetime coordinates.
Indeed while zero-momentum
standard cohomology in bosonic closed string theories does not
vanish at ghost number one, the $x_0$-cohomology does vanish [\astashkevich].
This is rather natural, as we expect no candidate anomalies in bosonic
string theory. In superstring theory, we will see that while there
are many zero-momentum semi-relative classes at $G=1$,
only one state survives, and only for type IIB or type I
string theory.

In an open-closed string theory, to lowest possible order in the 
topological genus expansion, the term  $\langle\Psi, D\rangle$ arises
from D-branes and orientifolds (once-punctured disks or crosscaps). If there
is a candidate $x_0$-cohomology class, explicit computation is necessary
to see that it does not appear in $\ket{D}$. This is precisely
what happens for type I string theory; the candidate class exists
and its disappearance from $\ket{D}$ fixes the gauge group to be
$SO(32)$. In a purely closed
string theory, one-point functions would arise should one-point amplitudes
for closed string states fail to vanish. This is not supposed to happen
for type IIB string theory, thus despite the existence of a candidate
class for an anomaly, the anomaly is not present.

\section{Prototype calculations}

In this section, 
we will discuss the two prototype calculations that are useful
for the computation of $x_0$-cohomology.  For these two cases, we will  
write the gauge transformations and find the gauge invariant states.
For this purpose, we need the description of bispinors in terms of 
differential forms. The well-known expansions are
$$\eqalign{
a_{\alpha\beta} &= a^{(1)}_{\mu_1}  \, (C\Gamma^{\mu_1})_{\alpha\beta} 
\,+ \,a^{(3)}_{\mu_1\mu_2\mu_3} \, (C\Gamma^{\mu_1\mu_2\mu_3})_{\alpha\beta} 
\, + \,a^{(5+)}_{\mu_1\cdots\mu_5}  
\, (C\Gamma^{\mu_1\cdots\mu_5}_+ )_{\alpha\beta} \, , \cr
b_{\alpha'\beta'} &= b^{(1)}_{\mu_1}  (C\Gamma^{\mu_1})_{\alpha'\beta'} 
+ b^{(3)}_{\mu_1\mu_2\mu_3}  (C\Gamma^{\mu_1\mu_2\mu_3})_{\alpha'\beta'} 
+ a^{(5-)}_{\mu_1\cdots\mu_5}  
(C\Gamma^{\mu_1\cdots\mu_5}_{-} )_{\alpha'\beta'} \,,  \cr
c_{\alpha\beta'} &= c^{(0)} \,\,(C)_{\alpha\beta'} 
\,+\, c_{\mu_1\mu_2} ^{(2)}\, (C\Gamma^{\mu_1\mu_2})_{\alpha\beta'}  \,+\,
c_{\mu_1\cdots\mu_4} ^{(4)}\,  
(C\Gamma^{\mu_1\cdots\mu_4})_{\alpha\beta'}\, .\cr}\eqn\exq$$
The spinor indices 
$\alpha$ and $\alpha'$ are identified with the irreducible spinor
representations $16$ and $16'$ respectively.  The degree of the differential
form is 
indicated 
by the superscript,
with the plus/minus in the five-forms indicating self dual
and anti-self 
dual pieces.  In the first two equations, 
the one and five forms appear
in the 
symmetric part of 
the product of spinor indices, while the three form appears in
the antisymmetric 
part of the product. The matrices $C_{\alpha\beta'}$ and $C_{\alpha'\beta}$ 
are the 
nonvanishing 
pieces of the $32\times 32$ charge conjugation matrix. As such,
they satisfy $\Gamma^{\mu\,T} C = - C \Gamma^\mu$.  The totally antisymmetric
products of  $\Gamma$ matrices satisfy a basic relation
$$\Gamma^{\mu_1\cdots \mu_n}\Gamma^\nu 
= \Gamma^{\mu_1\cdots \mu_n\nu}  
+ \{\delta^{\nu\mu_n} \Gamma^{\mu_1\cdots\mu_{n-1}} \pm \cdots \}\,. \eqn\mult$$

The gauge parameters that appear in the calculations of $x_0$-cohomology
are of the form $\Lambda_{\mu \, \alpha\beta}$, $\Lambda_{\mu \, \alpha'\beta'}$
or $\Lambda_{\mu \, \alpha\beta'}$;    bi-spinors with an additional
vector index. The spinor indices can be treated as in \exq. 
 The presence of the additional vector index allows for
two  operations. In the first one, that index can be antisymmetrized with
respect 
to the other antisymmetric indices. For example, doing this to the two-forms
$\Lambda^{(2)}_{\mu, \nu_1\nu_2}$ 
yields a three-form  $\Lambda^{(2)}_{[\mu, \nu_1
\nu_2]}$. We will denote the resulting three-form as $a\Lambda^{(2)}$, where
the prefix $a$ indicates taking the antisymmetric part.  In the
second operation, 
we can contract the vector index against one of the form
indices. For example $\Lambda^{(2)}_{\mu, \nu_1 \nu_2}$ would this
time yield the one form $\Lambda^{(2)}_{\mu, \mu\nu_1}$. This one form
will be denoted as 
$t\Lambda^{(2)}$, where the prefix $t$ indicates taking trace.
Although it is also possible to combine the indices to
form a partially-symmetric three-form
tensor, such a tensor will not appear in the cohomology calculations below.

\goodbreak
\medskip
\noindent
\underbar{Gauge System I}
~We now begin with the first prototype problem. This is the gauge system
described by a gauge parameter with two spinor indices of the same type.
We have 
$$\eqalign{
\delta \, a_{\alpha\beta'}  &= \,  \Lambda_{\mu\, \alpha\gamma} \,\,
\Gamma^{\mu\, \gamma}_{~~~\beta'} \, , \cr
\delta \, b_{\beta'\alpha}  &= \,  \Lambda_{\mu\, \gamma\alpha} \,\,
\Gamma^{\mu\, \gamma}_{~~~\beta'} \,  . \cr}\eqn\gtran$$
It is convenient to form linear combinations 
$$\delta \, {\cal A}_{\pm \alpha\beta'} 
\equiv\delta 
\, ( a_{\alpha\beta'} 
\pm b_{\beta'\alpha}   ) = \,\,  \Lambda_{\pm\mu\,\alpha\gamma} \,
\Gamma^{\mu\, \gamma}_{~~~\beta'}   \, , \eqn\gtrans$$
where $\Lambda_{+\mu}$ and $\Lambda_{-\mu} $ denote the symmetric 
and antisymmetric parts (with respect to the spinor indices) of
the gauge parameter $\Lambda_\mu$. 
The fields $ {\cal A}_{\pm \alpha\beta'} $ define
 zero-, two- and four-forms ${\cal A}_\pm^{(0)},
{\cal A}_\pm^{(2)} , $ and ${\cal A}_\pm^{(4)}$. 
The symmetric  gauge parameters
define  
one-forms $\Lambda^{(1)}_\mu$ and self-dual five forms $\Lambda^{(5+)}_\mu$,
while 
the antisymmetric gauge parameters define  three forms $\Lambda^{(3)}_\mu$.
Using the expansions \exq,  substituting into \gtrans, and using \mult\ we find
$$\delta  {\cal A}_+^{(0)}  \sim   t\Lambda^{(1)}\,, \quad
\delta {\cal A}_{+} ^{(2)} \sim   a\Lambda^{(1)} \,, \quad
\delta{\cal A}_{+}^{(4)}   \sim   t\Lambda^{(5+)}\,, \eqn\gtpluss$$
implying that all the ${\cal A}_+$ fields can be gauged away.
For the  ${\cal A}_-$ fields we find 
$$\delta  {\cal A}_-^{(0)}  \sim   0\,,\quad
\delta {\cal A}_{-} ^{(2)} \sim   t\Lambda^{(3)} \,, \quad
\delta{\cal A}_{-}^{(4)}   \sim   a\Lambda^{(3)} \, . \eqn\gtminus$$
We thus see that the scalar ${\cal A}_-^{(0)}$ cannot be gauged away. 
It corresponds 
to  ${\cal A}_{- \alpha\beta'}$ 
$ = C_{\alpha\beta'}$,  $ {\cal A}_{+ \alpha\beta'}$
$ = 0$, or, up to an overall irrelevant factor, to
$$a_{\alpha\beta'} = - b_{\beta'\alpha} = C_{\alpha\beta'}\, . \eqn\answone$$
In a nut-shell, the gauge system contained two scalars
and a single one-form $\Lambda^{(1)}_\mu$ gauge parameter whose trace could only
be used to gauge away one of the scalars.

\goodbreak
\medskip
\noindent
\underbar{Gauge System II}
We now have a gauge parameter of mixed spinor
type, and gauge transformations reading
$$\eqalign{
\delta \, a_{\alpha\beta}  &= \,  \Lambda_{\mu\, \alpha\gamma'} \,\,
\Gamma^{\mu\, \gamma'}_{~~~\beta} 
\, \,  = 
 (\Lambda_\mu \, \Gamma^\mu)_{\,\alpha\beta} \,,\cr 
\delta \, b_{\alpha'\beta'}  &= \,  \Lambda_{\mu\, \gamma\beta'} \,\,
\Gamma^{\mu\, \gamma}_{~~~\alpha'} \, 
=    ( \Gamma^{\mu\,T} \Lambda_\mu)_{\alpha'\beta'}  . \cr}\eqn\gtrano$$
We again 
expand the gauge parameters and the fields  in terms of differential forms
and  then find the following gauge transformations 
$$\eqalign{
\delta a^{(1)}  &\sim   \,a\Lambda^{(0)}  - t\Lambda^{(2)} \,,\quad\,\,\,
\delta a^{(3)} \sim  \, \,a\Lambda^{(2)}  - t\Lambda^{(4)} \,,\quad\,\,\,
\delta a^{(5+)}   \sim  \, a_+\Lambda^{(4)}\,,\cr
\delta b^{(1)}  &\sim   - a\Lambda^{(0)}  - t\Lambda^{(2)} \,,\quad
\delta b^{(3)} \sim   -a\Lambda^{(2)}  - t\Lambda^{(4)} \,,\quad
\delta b^{(5-)}   \sim   -a_-\Lambda^{(4)}\,, \cr } \eqn\gtplussa$$
where the subscripts in $a_\pm$ indicate taking the self-dual or 
anti-self-dual
combinations. 
Considering the gauge transformations 
of sums and differences of the $a$ and $b$ one forms, and of the three forms,
we see that they can all be gauged away. In addition, 
 the five-forms can be gauged
away 
separately.  All of $a_{\alpha\beta}$ and $b_{\alpha'\beta'}$ is pure gauge.

\section{$x_0$-cohomology in the $(-{1/ 2} , -{1/ 2}) $ picture}

We discuss in turn the IIB, type-I and IIA superstrings. 

\noindent 
\underbar{IIB superstring}.~~There is no  cohomology at $G=-1$, 
in fact, there are no zero-momentum 
$L_0 = \bar L_0= 0 $ states at this ghost number  because the 
candidate oscillators $\beta_0, \bar \beta_0, b_0, \bar b_0$ vanish on the
vacuum. This fact does not change
when we include polynomials in $x_0$.  At $G=0$  the set of physical
states is described 
by $f_{\alpha\beta} 
\ket{-\fracs12}^\alpha \otimes \ket{-\fracs12}^\beta$, where
$f_{\alpha\beta}$ is a constant bispinor, and both 
$\ket{-\fracs12}^\alpha$ and $\ket{-\fracs12}^\beta$ are Grassmann odd.  
Since there are no candidate $G=-1$ states, all of these states remain physical
states in $x_0$-cohomology. This is expected  since the physical states
represent constant field strengths for the RR gauge fields and,
as such, they are 
gauge invariant and cannot be gauged away.  
The $G=1$ cohomology requires some computation. The candidate states are
the ones found earlier
$$\Psi  \, = \,
  -a_{\alpha\beta'} \, \bar\gamma_0 \, \ket{-\fracs12}^\alpha 
\otimes \ket{-\fracs12}^{\beta'}
\, + \, b_{\beta' 
\alpha} \, \gamma_0  \ket{-\fracs12}^{\beta'} \otimes 
\ket{-\fracs12}^\alpha\, .\eqn\field$$
The gauge parameters are of the form
$$\Lambda  = 
x_0^\mu \, \Lambda_{\mu\,\alpha\beta} \ket{-\fracs12}^\alpha 
\otimes \ket{-\fracs12}^\beta\, .
\eqn\gpar$$
We do not include a term quadratic in $x_0$ since, upon action by $Q$, it 
would give rise to a term containing $c_0^+$  and there are no such terms 
in  \field. Since $Q$ reduces the powers of $x_0$ by one or two units, higher
polynomials in $x_0$ are not necessary.  Recalling that  
$\ket{-\fracs12}^\alpha$ and 
$\ket{-\fracs12}^{\beta'}$
are respectively Grassmann odd and Grassmann even, a short computation gives
gauge transformations identical to those in \gtran. 
As discussed there, we have a gauge invariant state
in the cohomology. Making use of  \answone\ and \field,
 the state is found to be
$$\ket{ {\cal A}^0 } = \, 
C_{\alpha\beta'} \, (  \bar\gamma_0 \, \ket{-\fracs12}^\alpha 
\otimes \ket{-\fracs12}^{\beta'}
\, + \, \gamma_0  \ket{-\fracs12}^{\beta'} 
\otimes \ket{-\fracs12}^\alpha ) \, .\eqn\fieldr$$
This semirelative $x_0$-cohomology class at ghost number one represents
a candidate anomaly, as discussed earlier.  
This completes our computation of
$x_0$-cohomology 
in the $(-{1/ 2} , -{1/ 2})$ picture for the IIB string.

\medskip
\noindent
\underbar{Type I superstring}.~ We now consider  
the Type I superstring, whose BRST complex
is  defined as the subcomplex of the IIB superstring complex spanned by 
states that are preserved by an exchange
of  left and right-movers. 
More precisely, 
when the left and right movers are in the same picture, there is an
obvious way to map the left moving and right moving state spaces into each 
other, namely, we exchange holomorphic and antiholomorphic labels
on all oscillators and vacua.  
If we denote this map by ${\cal I}$, 
type-I states $\ket{\Psi}$ must satisfy 
$$T(\ket{\Psi})= \ket{\Psi}\,,\eqn\contran$$
where the exchange map $T$ is defined as  
$$ T \Bigl( a_{ij} \ket{i}\otimes \ket{j}\Bigr) = 
(-)^{ij} \, a_{ij} \,\,{\cal I}(\ket{j})\otimes {\cal I}(\ket{i})\,,
\eqn\typeone$$
and the sign prefactor takes into account the grassmanality of the
states that have been exchanged. As in the
IIB case, there is no $G=-1$ cohomology since there are no
candidate states. At $G=0$, the cohomology consists of the states
$f_{\alpha\beta} 
\ket{-\fracs12}^\alpha \otimes \ket{-\fracs12}^\beta$, where
$f_{\alpha\beta}=
-f_{\beta\alpha}$ is a constant anti-symmetric bispinor field.
Note that $f_{\alpha\beta}$ is anti-symmetric on account
of \typeone\ given that $\ket{-\fracs12}^\alpha$
and $\ket{-\fracs12}^\beta$ are both Grassman odd. 

At $G=1$, the candidate states are those of \field\ where
$a_{\alpha\beta'}
=-b_{\beta'\alpha}$ (since  
$\ket{-\fracs12}^\alpha$
and $\ket{-\fracs12}^{\beta'}$ commute). One can compute the cohomology using
Gauge System I remembering that the gauge parameter
$\Lambda$ must be
anti-symmetric in its spinor indices and that $a_{\alpha\beta'}$=  
$-b_{\beta'\alpha}$. Therefore, equation \gtpluss\ is unnecessary while
equation \gtminus\ is the same as in the Type IIB case. So the
$G=1$ cohomology of the Type I superstring contains the same scalar
as in \fieldr. 

\medskip
\noindent
\underbar{IIA superstring}.
There is  no $x_0$-cohomology for $G=-1$, and 
at $G=0$ the semirelative 
cohomology classes $f_{\alpha\beta'} 
\ket{-\fracs12}^\alpha\otimes \ket{-\fracs12}^{\beta'}$ remain 
nontrivial in $x_0$- cohomology. For $G=1$, 
the candidate states are the semirelative classes
$$\Psi  \, = \,  
-a_{\alpha\beta} \, \bar\gamma_0 \, \ket{-\fracs12}^\alpha \otimes 
\ket{-\fracs12}^{\beta}
\, + \, b_{\alpha'\beta' } 
\, \gamma_0  \ket{-\fracs12}^{\alpha'} \otimes 
\ket{-\fracs12}^{\beta'}\,.\eqn\fieldo$$
The gauge parameters are taken to be  of the form
$$\Lambda  = x_0^\mu \, 
\Lambda_{\mu\,\alpha\beta'} \ket{-\fracs12}^\alpha \otimes 
\ket{-\fracs12}^{\beta'}\, .\eqn\gparo$$
A little calculation gives the gauge transformations
considered in \gtrano. It follows from the
analysis of these equations that  the  IIA $x_0$-cohomology at $G=1$ is absent.

\section{Finite $x_0$-cohomology in the $(-{3/ 2} , -{1/ 2})$ picture}

\medskip 
\noindent
\underbar{IIB superstring}
Consider IIB theory at $G=-1$.  The semirelative  classes found earlier are 
$$\Psi  \, = \,  a_{\alpha\beta} \,\beta_0 \, 
\ket{-\fracs32}^\alpha \otimes \ket{-\fracs12}^{\beta}
\, -\fracs12  \, b_{\alpha'\beta' } \, \beta_0^2\bar\gamma_0  
\ket{-\fracs32}^{\alpha'} \otimes \ket{-\fracs12}^{\beta'}\,.\eqn\fieldox$$
Here the ket  $\ket{-\fracs32}^\alpha \otimes \ket{-\fracs12}^{\beta}$ 
is Grassmann odd, with
$\ket{-\fracs32}^\alpha$  even and $\ket{-\fracs12}^\beta$ odd.  
It is now sufficient to consider
the  $x_0$-dependent
$G=-2$ gauge parameter
$$\Lambda =  \fracs12 \,  x_0^\mu \,\Lambda_{\mu\,\alpha'\beta} 
 \beta_0^2 \ket{-\fracs32}^{\alpha'} \otimes \ket{-\fracs12}^{\beta}\, , 
\eqn\dblazy$$
and the resulting gauge transformations read
$$\eqalign{
\delta \, a_{\alpha\beta}  &= \,  \Lambda_{\mu\, \gamma'\beta} \,\,
\Gamma^{\mu\, \gamma'}_{~~~\alpha}  =  
\,(\Gamma^{\mu\, T} \Lambda_\mu)_{\alpha\beta}\, \,,  \cr
\delta \, b_{\alpha'\beta'}  &= \, \,  \Lambda_{\mu\, \alpha'\gamma} \,\,
\Gamma^{\mu\, \gamma}_{~~~\beta'}  
= (\Lambda_\mu \Gamma^\mu)_{\alpha'\beta'}\,  . \cr}\eqn\gtranuzz$$
which 
after  
the exchange of primed and unprimed indices and the exchange of $a$ and $b$,
coincide precisely with those considered in \gtrano.  We therefore 
conclude that 
there is no $x_0$-cohomology at $G=-1$ for IIB string theory.

\medskip

Let us now consider cohomology at $G=0$.  The candidate states are
$$\Psi  \, = \,  a_{\alpha'\beta} \, 
\ket{-\fracs32}^{\alpha'} \otimes \ket{-\fracs12}^{\beta}
\,+ \, b_{\alpha\beta' } \, \beta_0\bar\gamma_0  
\ket{-\fracs32}^{\alpha} \otimes \ket{-\fracs12}^{\beta'}\,.\eqn\fieldoxt$$
Here the ket  $\ket{-\fracs32}^{\alpha'} \otimes \ket{-\fracs12}^{\beta}$ 
is Grassmann even, with
both $\ket{-\fracs32}^{\alpha'}$ and $\ket{-\fracs12}^{\beta}$  odd.  
As we  will claim  that there
is a gauge invariant scalar, it is  necessary to consider the most
general  $x_0$-dependent  $G=-2$ gauge parameter.  We take
$$\eqalign{
\Lambda  &=  x_0^\mu \,\Bigl( \Lambda_{\mu\,\alpha\beta} \beta_0
 \ket{-\fracs32}^{\alpha} \otimes \ket{-\fracs12}^{\beta}
+\fracs12 \widetilde\Lambda_{\mu\,\alpha'\beta'}  \beta_0^2 \bar\gamma_0  
\ket{-\fracs32}^{\alpha'} \otimes \ket{-\fracs12}^{\beta'} \Bigr) \cr
&~~~+\,\fracs12\,
\Omega_{\alpha'\beta}\, c_0^+ \,\beta_0^2\, 
\ket{-\fracs32}^{\alpha'} \otimes \ket{-\fracs12}^{\beta}\,,\cr}\eqn\candg$$
where the last term,  necessary in order to cancel terms containing a
$\beta_0^2\bar\gamma_0^2$ factor in the gauge variations, fixes  
$\Omega_{\alpha'\beta} = - \widetilde\Lambda_{\mu\, \alpha'\gamma'}
\Gamma^{\mu\,\gamma'}_{~~~\beta}$. The  gauge transformations  then read
$$\eqalign{
\delta \, a_{\alpha'\beta}  &= \,  \Lambda_{\mu\, \gamma\beta} \,\,
\Gamma^{\mu\, \gamma}_{~~~\alpha'} \, \, 
+ \widetilde\Lambda_{\mu\, \alpha'\gamma'} \,\,
\Gamma^{\mu\, \gamma'}_{~~~\beta} \, ,\cr
\delta \, b_{\alpha\beta'}  &= \, \,  \Lambda_{\mu\, \alpha\gamma} \,\,
\Gamma^{\mu\, \gamma}_{~~~\beta'} 
+  \widetilde\Lambda_{\mu\, \gamma'\beta'} \,\,
\Gamma^{\mu\, \gamma'}_{~~~\alpha} \,  . \cr}\eqn\gtranuqw$$
This is essentially gauge system I  (\gtran) with two gauge parameters
$\Lambda$ and
$\widetilde\Lambda$ 
playing similar roles. We now confirm that  there is a gauge 
invariant scalar.  To show this,
it is convenient to rewrite the above equations as
$$\eqalign{
\delta \, a_{\alpha'\beta}  &=\bigl (  \Gamma^{\mu \,T} \, \Lambda_{\mu} 
\, \, + \widetilde\Lambda_\mu\,
\Gamma^{\mu}  \bigr)_{\alpha'\beta} \cr
\delta \, b_{\beta\alpha'}  &= \bigl( \Gamma^{\mu \,T} \, \Lambda_{\mu} ^T
\, +  \,\widetilde\Lambda_{\mu}^T  \Gamma^{\mu} \,\bigr)_{\alpha'\beta}
\,  . \cr}\eqn\gtranuc$$
Defining linear combinations, we now find
$$\delta {\cal A}_{\pm \alpha'\beta} \equiv \delta (a_{\alpha'\beta} 
\pm b_{\beta\alpha'})
= \bigl (  \Gamma^{\mu \,T} \, \Lambda_{\pm\mu} 
\, \, + \widetilde\Lambda_{\pm\mu}\,
\Gamma^{\mu}  \bigr)_{\alpha'\beta}\eqn\zmm $$
where the $\pm$ subscripts in the gauge parameters indicate the 
exchange property
for spinor indices. Since the gauge transformations of ${\cal A}_-$ 
involve the
antisymmetric gauge parameters, and they only contain three forms, it is clear 
that the scalar part of ${\cal A}_-$ cannot be gauged away.
We thus have ${\cal A}_{-\alpha'\beta} = C_{\alpha'\beta} \,, 
\,  {\cal A}_{+\alpha'\beta} = 0$
and the state representing the cohomology class is given by
$$\ket{{\cal A}_-} = C_{\alpha'\beta} 
\Bigl( \ket{-\fracs32}^{\alpha'}\otimes\ket{-\fracs12}^\beta
- \beta_0 \bar\gamma_0 \ket{-\fracs32}^\beta\otimes \ket{-\fracs12}^{\alpha'} 
\Bigr)\,.\eqn\zma$$
This state is the zero momentum axion.  
Just as the zero momentum ghost-dilaton,
it would be trivial in absolute cohomology.

We now consider $G=1$ cohomology.  
The candidate states and  gauge parameters  read
$$\eqalign{
\Psi  \,\,&= \, f_{\alpha'\beta'} \, \bar\gamma_0 \ket{-\fracs32}^{\alpha'} 
\otimes \ket{-\fracs12}^{\beta'}\,,\cr
\Lambda  \,\, &= \, \
x_0^\mu\, \Lambda_{\mu\alpha'\beta} \ket{-\fracs32}^{\alpha'} 
\otimes \ket{-\fracs12}^\beta\, .\cr} \eqn\cohcom$$
The resulting gauge transformations
$\delta f_{\alpha'\beta'} = - \Lambda_{\mu\, \alpha'\gamma} 
\Gamma^{\mu\, \gamma}_{~~~\beta'}$, are a subset of those in gauge system II.  
It then follows that  there is no
IIB  $x_0$-cohomology at $G=1$.

\medskip 
\noindent
\underbar{Type I superstring}.~
Since the left-moving and right-moving pictures are different, the
Type I string fields $\ket{\Psi}$ will be defined to satisfy the condition
$$ T ( \overline{\cal Y}_0 \ket{\Psi}) =
 \overline{\cal Y}_0 \ket{\Psi}\,,\eqn\type$$
where $T$ is defined in \typeone\ and
$ \overline{\cal Y}_0 $ is defined in \itiss.  Note that the state
$\overline{\cal Y}_0 \ket{\Psi}$ is in the diagonal $(-3/2,-3/2)$ picture,
and
thus the action of $T$ is well defined.

First consider the
$G=-1$ cohomology. If $\Psi$ is defined as in \fieldox, condition
\type\ implies that $a_{\alpha\beta}=
a_{\beta\alpha}$ and
$b_{\alpha'\beta'}=
b_{\beta'\alpha'}$. Defining the gauge parameter 
$$\eqalign{
\Lambda  &=  \fracs12 \,  x_0^\mu \,\Lambda_{\mu\,\alpha'\beta}\,\,
( \beta_0^2 \ket{-\fracs32}^{\alpha'} \otimes \ket{-\fracs12}^{\beta}
-\fracs13  \,\beta_0^3 \,\bar\gamma_0 \,\ket{-\fracs32}^{\beta} \otimes
\ket{-\fracs12}^{\alpha'} )\cr
&\,\,- \fracs16 \,c_0^+ \,\beta_0^3 \,\Lambda_{\mu\,\alpha'\beta}\,
\Gamma^{\mu\, \alpha'}_{~~~\alpha} 
\ket{-\fracs32}^{\beta} \otimes
\ket{-\fracs12}^{\alpha} 
  \, ,\cr} \eqn\gtype$$
which can be verified to satisfy \type,
one finds the  gauge transformations as \gtranuzz\ for the
symmetric parts of the fields.  
It follows that there is no cohomology.

At $G=0$, the string field $\Psi$ of \fieldoxt\ satisfies \type\ if
$a_{\alpha'\beta}=
b_{\beta\alpha'}$. Similarly, the gauge parameter of \candg\
satisfies \type\ if
$\Lambda_{\mu\,\alpha\beta}=
\Lambda_{\mu\,\beta\alpha}$, and
$\tilde\Lambda_{\mu\,\alpha'\beta'}=
\tilde\Lambda_{\mu\,\beta'\alpha'}$.
So the gauge transformation of ${\cal A}_{+\alpha'\beta}$ is the same as in
\zmm, while the field
${\cal A}_{-\alpha'\beta}$ vanishes. Therefore, there is no cohomology at 
$G=0$.

At $G=1$ the candidate state is the same as in \cohcom, but with
$f_{\alpha'\beta'}= - 
f_{\beta'\alpha'}$. The appropriate gauge parameter in the type I complex
is
$$\eqalign{
\Lambda  \,\, &= \, x_0^\mu \,\Lambda_{\mu\alpha'\beta} \,\,
(\, \ket{-\fracs32}^{\alpha'}
\otimes \ket{-\fracs12}^\beta
+\beta_0\bar\gamma_0
\ket{-\fracs32}^{\beta}
\otimes \ket{-\fracs12}^{\alpha'}\, )\cr
&\,\,\,\, +  c_0^+ \,\Lambda_{\mu\alpha'\beta}\, 
\Gamma^{\mu\alpha'}_{~~~\alpha}
\, \b_0 \,\ket{-\fracs32}^{\beta}
\otimes \ket{-\fracs12}^{\alpha}\,  \, .\cr}
 \eqn\cohcomtwo$$
The resulting gauge transformation is
$\delta f_{\alpha'\beta'} = - \Lambda_{\mu\, \alpha'\gamma}
\Gamma^{\mu\, \gamma}_{~~~\beta'} + 
\Lambda_{\mu\, \beta'\gamma}
\Gamma^{\mu\, \gamma}_{~~~\alpha'}$, and therefore, using the same
argument as in the Type IIB case, we conclude that
there is no cohomology at $G=1$.

\medskip
\noindent
\underbar{IIA superstring}.~We begin with the case $G=-1$. 
Here the candidate states are
$$\Psi = a_{\alpha'\beta} \, \beta_0 \ket{-\fracs32}^{\alpha'} 
\otimes\ket{-\fracs12}^\beta - \fracs12 
\, b_{\alpha\beta'}
\, \beta_0^2 \bar\gamma_0 \ket{-\fracs32}^\alpha\otimes\ket{-\fracs12}^{\beta'} 
\, .\eqn\twoago$$
After consideration of the appropriate gauge parameters, one finds gauge 
transformations  identical to those in \gtranuqw.  We therefore conclude
that  one scalar survives in $x_0$-cohomology.
This is the state 
$$ \ket{{\cal G}} = C_{\alpha'\beta} 
\Bigl(   \beta_0 \ket{-\fracs32}^{\alpha'} \otimes\ket{-\fracs12}^\beta +
 \fracs12  \,\, \beta_0^2 \bar\gamma_0 
\ket{-\fracs32}^\beta\otimes\ket{-\fracs12}^{\alpha'} \Bigr)\,, \eqn\newst$$
which, being at $G=-1$ represents a symmetry generator. The associated
scalar charge can 
be identified 
with RR-charge, or as the momentum generator along an extra dimension
curled up into a circle.
This extra dimension is the eleventh direction of  M-theory.  
The fact that no extra
states 
were found for IIB superstrings is consistent with the viewpoint that the 
extra dimensions in  F-theory seem to be nondynamical.

At  $G=0$, 
the candidate states read
$$\Psi = 
a_{\alpha\beta} \ket{-\fracs32}^\alpha\otimes
\ket{-\fracs12}^\beta + b_{\alpha'\beta'} \,\beta_0
\bar\gamma_0 \, \ket{-\fracs32}^{\alpha'}\otimes\ket{-\fracs12}^{\beta'}
\,.  \eqn\csas$$
The gauge 
transformations for this case are found to be  those in \gtranuzz. It thus
follows that the IIA $x_0$-cohomology at $G=0$ vanishes.  Finally, for $G=1$ we
find that all the candidate states $f_{\alpha\beta'} \bar\gamma_0
\ket{-\fracs32}^\alpha \otimes\ket{-\fracs12}^{\beta'}$ 
can be gauged away in $x_0$-cohomology. This completes out calculations of 
$x_0$ -cohomology, the results of which 
are summarized in the table below.

\vskip .1in
$$\hbox{\vbox{\offinterlineskip
\def\strut{\hbox{\vrule height 25pt depth 20pt width 0pt}}
\hrule
\halign{
\strut\vrule#\tabskip 0.1in&
\hfil$#$\hfil &
\vrule#&
\hfil$#$\hfil &
\vrule#&
\hfil$#$\hfil &
\vrule#&
\hfil$#$\hfil &
\vrule#&
\hfil$#$\hfil &
\vrule#\tabskip 0.0in\cr
& G.N.&& \hbox{IIA}\,[-{3\over 2}\, , - {1\over 2}~  ] 
&& {\hbox{IIB}\atop \hbox{Type I}}\,[-{3\over 2}\, , - {1\over 2}~  ]
&& \hbox{IIA}\,\,[-{1\over 2}\, , - {1\over 2}~  ]
&&  {\hbox{IIB}\atop \hbox{Type I}}\, [-{1\over 2}\, , - {1\over 2}~  ]
& \cr\noalign{\hrule}
&-1 
&& \matrix{1 \, (\sim \partial/\partial x^{11})
\cr (\hbox{see}\, \newst) }
&& \matrix{0\hskip-6pt/\cr 0\hskip-6pt/}
&& 0\hskip-6pt/
&& \matrix{0\hskip-6pt/\cr 0\hskip-6pt/} & \cr\noalign{\hrule}
& 0
&& 0\hskip-6pt/
&& \matrix{1 \,\,(\hbox{axion},\zma)\cr 0\hskip-6pt/}
&&  \matrix{ 256 \cr  (\hbox{field strengths})}
&& \matrix{ 256 \cr 120}\,\, (\hbox{field strengths})  & 
\cr\noalign{\hrule}
& +1
&& 0\hskip-6pt/
&& 0\hskip-6pt/
&& 0\hskip-6pt/
&& \matrix{ 1\cr 1}  \,\,(\hbox{anomaly}, \fieldr) 
&\cr\noalign{\hrule}
 }}}$$
{\bf Table.} Summary of $x_0$-semirelative cohomology classes for
IIA, IIB and type I superstrings. 
\medskip
\medskip

\vskip 20pt
\centerline{\bf Acknowledgements} 
We would like to thank A. Belopolsky for conversations
on picture changing operators, and S. Rangoolam for conversations on
finite/infinite cohomology. We are grateful
to Harvard University for their hospitality.  
\refout

\end